\documentclass[aps,prc,showkeys,reprint,superscriptaddress,nofootinbib,amsmath,amssymb,floatfix,longbibliography]{revtex4-2}

\usepackage{bm}
\usepackage{booktabs}
\usepackage{tikz}
\usepackage[colorlinks=true,allcolors=blue]{hyperref}
\usepackage{graphicx}
\usepackage{epstopdf}

\usepackage[normalem]{ulem}

\newcommand{\ad}[1]{\alpha^{\dagger}_{#1}}
\newcommand{\an}[1]{\alpha_{#1}}
\newcommand{\vac}{|\widetilde{0}\rangle}
\newcommand{\bra}{\langle\widetilde{0}|}
\newcommand{\vb}{\bar v}

\newcommand{\Amat}{\mathcal{A}}
\newcommand{\Bmat}{\mathcal{B}}

\begin{document}

\title{Second quasiparticle random-phase approximation:
formalism for open-shell nuclei}

\author{Deni Vale}
\email[]{dvale@iv.hr}
\affiliation{Istrian University of Applied Sciences, Preradovićeva 9D, HR-52100 Pula, Croatia}
\affiliation{Faculty of Engineering, Juraj Dobrila University of Pula, Alda Negrija 6, HR-52100 Pula, Croatia}

\date{\today}

\begin{abstract}
We derive the equations of the second quasiparticle random-phase approximation
(SQRPA), the extension of the quasiparticle RPA (QRPA) in which the excitation
operator contains four-quasiparticle components in addition to the usual
two-quasiparticle ones, so that pairing correlations and two-particle--two-hole-like
fragmentation are treated on the same footing.  Starting from the
Hartree--Fock--Bogoliubov (HFB) vacuum and using the equation-of-motion method in the
quasiboson approximation, we give the complete set of vertices
$H^{40}$, $H^{31}$ and $H^{22}$ in terms of the antisymmetrized interaction and the
Bogoliubov amplitudes $U$ and $V$, and explicit expressions for all blocks of
the SQRPA matrices. 
The matrix $\mathcal B$ is shown to be nonzero only in the
two-quasiparticle sector, and the coupling between the two sectors is
governed by $H^{31}$. Elimination of the four-quasiparticle space yields an
energy-dependent QRPA problem in which its subtracted version (SSQRPA) is formulated. The SSQRPA reduces to the QRPA at zero energy and preserves the zero-energy QRPA solution. Without pairing the equations reduce
to the standard second RPA (SRPA), and all expressions are verified to machine precision in an independent Fock-space
calculation.  In an exactly solvable pairing model the SQRPA finds the correct states but lowers their energies below
the QRPA, and the spurious mode of broken particle-number symmetry becomes unstable.  The subtraction restores this mode and reduces the shift, but for six and eight levels neither version improves on the
QRPA for the lowest excitation. A projected configuration interaction (PCI) in the same configuration space reproduces the low-lying exact spectrum with good accuracy, indicating that the SQRPA deviations arise primarily from the approximations used to treat this space rather than from its limited size. Possible sources include the quasiboson approximation, the uncorrelated reference state, and the breaking of particle-number symmetry.  For larger systems the shifts relative to the QRPA decrease roughly as the inverse number of levels, and for twenty
levels, where the QRPA overestimates the exact excitation energy, the SSQRPA is closer to it. The formalism in this work therefore provides a starting point for second-order quasiparticle calculations with realistic interactions.
\end{abstract}

\keywords{second quasiparticle random-phase approximation, subtraction method, four-quasiparticle configurations, Hartree--Fock--Bogoliubov, particle-number projection,
picket-fence pairing model, open-shell nuclei}

\maketitle

%=====================================================================
\section{INTRODUCTION}
\label{sec:intro}
%=====================================================================

The random-phase approximation (RPA) and its quasiparticle version (QRPA) are the
standard tools for describing collective excitations and giant resonances of
medium-mass and heavy nuclei \cite{RingSchuck1980,BlaizotRipka1986}.  Their main limitation is
the restriction of the excitation space to one-particle--one-hole (or
two-quasiparticle) configurations, which gives the centroids and total strengths
of giant resonances but not their spreading width or fine structure.

The second RPA (SRPA) lifts this restriction by adding two-particle--two-hole
($2p2h$) configurations \cite{Rowe1986,Yannouleas,Yannouleas1983}.  Its equations can be obtained
by the equation-of-motion method, by the small-amplitude limit of time-dependent
density-matrix theory \cite{Yannouleas, Papakonstantinou2010}, and by a variational procedure due to da~Provid\^encia
\cite{Providencia}.  
%For many years realistic SRPA calculations were limited
%by their cost and by the systematic downward shift of the strength. 
For many years, realistic SRPA calculations were constrained by high computational demands.
Moreover, from a theoretical standpoint, applications were hindered by a systematic downward shift of the transition strength and instabilities in the lower part of excitation energy spectrum.
Both issues originate from the double counting of second-order correlations when employing effective residual interactions.
The subtraction method introduced by Tselyaev \cite{Tselyaev2007,Tselyaev2013}, which was later applied to the SRPA in
Ref.~\cite{Gambacurta2015}, removes these problems and has been used for
dipole, quadrupole and charge-exchange responses
\cite{GAMBACURTA2018163,Gambacurta2020, Yang2021, Yang2024}. Recently, a relativistic extension of the SRPA was introduced and applied to monopole and quadrupole giant resonances \cite{Vale2025, Vale2025RSRPA}.

All of these applications concern, or are restricted to, closed-shell nuclei.  The
generalization to superfluid nuclei requires the excitation operator to be
expressed in terms of quasiparticles and to contain four-quasiparticle
terms, i.e. the second quasiparticle RPA (SQRPA). This generalization is a recognized open problem, which a recent review of the subtracted SRPA lists among the main
theoretical extensions still to be made~\cite{GAMBACURTA2026104251}.  Very recently, Yang \textit{et al.}~\cite{Yang2026}
reported a self-consistent subtracted SQRPA (SSQRPA) based on HF+BCS with a Skyrme functional and applied it to the dipole
response of neutron-rich oxygen isotopes.  In that work the four-quasiparticle configurations shift the giant dipole
strength toward the observed energies and substantially reduce the deviation of the low-lying dipole states from
experiment.  To keep the calculation tractable, the four-quasiparticle space was restricted to configurations of
two-particle--two-hole type, defined by an occupation cutoff, and the pairing interaction was not included in the
four-quasiparticle sector.  The present work is complementary.  It provides the general formalism without restriction of
the four-quasiparticle space and tests it against an exact solution.

In this paper we present the SQRPA equations for a general Bogoliubov vacuum.  We use the equation-of-motion method
with the quasiboson approximation (QBA) taken with respect to the HFB vacuum.  The Hamiltonian is written in normal
order with respect to this vacuum, which gives the quasiparticle vertices that determine all matrix elements.  We
derive the matrices $\Amat$ and $\Bmat$ explicitly, show which vertices enter which blocks, and eliminate the
four-quasiparticle space to obtain an energy-dependent QRPA to which the subtraction method can be applied.  Without
pairing the equations reduce to the standard SRPA.  The algebra is verified in an independent Fock-space calculation.
We then apply the equations to a pairing model with an exact solution, the so-called picket-fence model, for which the matrices
can be computed both in the full Fock space and, for larger systems, in a compact representation of the paired
states.  To separate the effect of the approximations from that of the broken particle-number symmetry, we also
project the results onto good particle number.

The paper is organized as follows.  Section~\ref{sec:HQP} gives the
quasiparticle form of the Hamiltonian and the vertices.  Section~\ref{sec:SQRPA}
derives the SQRPA equations and the explicit matrices.
Section~\ref{sec:red} discusses the elimination of the four-quasiparticle space and the
subtraction.  
Sec.~\ref{sec:toy} applies the equations to a solvable pairing model, and
Sec.~\ref{sec:sum} gives conclusions.
Derivation details are given in the Appendices.

%=====================================================================
\section{HAMILTONIAN IN THE QUASIPARTICLE BASIS}
\label{sec:HQP}
%=====================================================================

\subsection{Bogoliubov transformation}

Let $a_r^\dagger$, $a_r$ be fermion operators.  Quasiparticle operators are
defined by the Bogoliubov--Valatin transformation \cite{Valatin1958, Bogoljubov1958}
\begin{align}
 \alpha_k &= \sum_r\left(U^*_{rk}\,a_r+V^*_{rk}\,a^\dagger_r\right),\nonumber\\
 \alpha^\dagger_k &= \sum_r\left(U_{rk}\,a^\dagger_r+V_{rk}\,a_r\right),
\label{eq:bog}
\end{align}
where the unitarity of the matrix $\bigl(\begin{smallmatrix}U&V^*\\V&U^*\end{smallmatrix}\bigr)$
ensures that the operators $\alpha_k$ obey the fermionic anticommutation relations, $\{\alpha_k,\alpha^\dagger_l\}=\delta_{kl}$, $\{\alpha_k,\alpha_l\}=0$.  The inverse is
\begin{align}
 a_r &= \sum_k\left(U_{rk}\,\alpha_k+V^*_{rk}\,\alpha^\dagger_k\right),\nonumber\\
 a^\dagger_r &= \sum_k\left(U^*_{rk}\,\alpha^\dagger_k+V_{rk}\,\alpha_k\right).
\label{eq:invbog}
\end{align}
%The HFB vacuum $\vac$ satisfies $\alpha_k\vac=0$ for all $k$.  Its densities are
%$\rho=V^*V^T$ and $\kappa=V^*U^T$.
The HFB vacuum $\vac$ satisfies $\alpha_k\vac=0$ for all $k$.
Its normal and pairing densities, $\rho_{rr'}=\langle\widetilde0|a^\dagger_{r'}a_r|\widetilde0\rangle$ and
$\kappa_{rr'}=\langle\widetilde0|a_{r'}a_r|\widetilde0\rangle$, are given in terms of the Bogoliubov matrices by
$\rho=V^*V^T$ and $\kappa=V^*U^T$ (see Appendix~\ref{app:hfb}).

\subsection{Normal-ordered Hamiltonian}

We consider an $A$-body system of fermions interacting through a two-body interaction, described by the Hamiltonian
\begin{equation}
 H=\sum_{rr'}\tilde t_{rr'}\,a^\dagger_ra_{r'}
 +\tfrac14\sum_{rsr's'}\vb_{rs,r's'}\,a^\dagger_ra^\dagger_sa_{s'}a_{r'},
\label{eq:H}
\end{equation}
where $t$ is the one-body kinetic-energy operator, $\mu$ is the chemical potential that fixes the mean particle number, and
$\tilde t=t-\mu$ (here $\mu$ multiplies the unit matrix). The quantity
$\vb_{rs,r's'}=\langle rs|v|r's'\rangle-\langle rs|v|s'r'\rangle$ is the antisymmetrized matrix element of the two-body interaction $v$ \cite{RingSchuck1980, Suhonen2007}.
It satisfies $\vb_{rs,r's'}=-\vb_{sr,r's'}=-\vb_{rs,s'r'}$ and $\vb_{rs,r's'}=\vb^*_{r's',rs}$.
Substituting Eq.~\eqref{eq:invbog} and normal-ordering with respect to $\vac$,
\begin{align}
 H&=E_0+H^{11}+\bigl(H^{20}+\mathrm{h.c.}\bigr)+H^{22}\nonumber\\
 &\quad+\bigl(H^{31}+\mathrm{h.c.}\bigr)+\bigl(H^{40}+\mathrm{h.c.}\bigr),
\label{eq:Hdec}
\end{align}
where the terms are given by
\begin{align}
 H^{11}&=\sum_{kl}e_{kl}\,\ad k\an l,\\
 H^{20}&=\tfrac12\sum_{k_1k_2}H^{20}_{k_1k_2}\,\ad{k_1}\ad{k_2},\\
 H^{22}&=\tfrac14\sum_{k_1k_2k_3k_4}H^{22}_{k_1k_2k_3k_4}\,\ad{k_1}\ad{k_2}\an{k_4}\an{k_3},\\
 H^{31}&=\tfrac16\sum_{k_1k_2k_3k_4}H^{31}_{k_1k_2k_3k_4}\,\ad{k_1}\ad{k_2}\ad{k_3}\an{k_4},\\
 H^{40}&=\tfrac1{24}\sum_{k_1k_2k_3k_4}H^{40}_{k_1k_2k_3k_4}\,\ad{k_1}\ad{k_2}\ad{k_3}\ad{k_4},
\end{align}
and each coefficient is antisymmetric in the creation indices and in the annihilation indices
separately.  Equivalently
\begin{align}
 H^{40}_{k_1k_2k_3k_4}&=\bra\an{k_4}\an{k_3}\an{k_2}\an{k_1}H\vac,\nonumber\\
 H^{31}_{k_1k_2k_3k_4}&=\bra\an{k_3}\an{k_2}\an{k_1}H\,\ad{k_4}\vac,
\label{eq:Hdef}
\end{align}
and $H^{22}$ is the connected two-body part of $\bra\an{k_2}\an{k_1}H\ad{k_3}\ad{k_4}\vac$, i.e. the contribution in which both quasiparticles interact at the same vertex.
The HFB condition reads $H^{20}=0$. The one-body part $e$, given explicitly in Eq.~\eqref{eq:ef} of
Appendix~\ref{app:vert}, is Hermitian for any Bogoliubov transformation, because $h$ is Hermitian and $\Delta$ is
antisymmetric.  Since the vacuum $\vac$ does not change under $\alpha_k\to\sum_lw_{kl}\alpha_l$ with $w$ unitary, the
quasiparticle basis can be chosen so that $e_{kl}=E_k\delta_{kl}$, where $E_k$ are the quasiparticle energies, positive
for the HFB solutions considered here.

\subsection{Vertices}

With $T$, $Y$, $X$ and $X^{a,b,c}$ defined below,
\begin{align}
 T_{k_1k_2k_3k_4}&=\sum_{rsr's'}\vb_{rs,r's'}\,U^*_{rk_1}U^*_{sk_2}V^*_{s'k_3}V^*_{r'k_4},\nonumber\\
 Y_{k_1k_2k_3;k_4}&=\sum_{rsr's'}\vb_{rs,r's'}\,U^*_{rk_1}V^*_{s'k_2}V^*_{r'k_3}V_{sk_4},\nonumber\\
 X_{k_1k_2k_3;k_4}&=\sum_{rsr's'}\vb_{rs,r's'}\,U^*_{rk_1}U^*_{sk_2}V^*_{r'k_3}U_{s'k_4},
\label{eq:TXY}
\end{align}
and
\begin{align}
 X^a_{k_1k_2k_3k_4}&=\sum_{rsr's'}\vb_{rs,r's'}\,U^*_{rk_1}U^*_{sk_2}U_{r'k_3}U_{s'k_4},\nonumber\\
 X^b_{k_1k_2k_3k_4}&=\sum_{rsr's'}\vb_{rs,r's'}\,V^*_{s'k_1}V^*_{r'k_2}V_{sk_3}V_{rk_4},\nonumber\\
 X^c_{k_1k_2k_3k_4}&=\sum_{rsr's'}\vb_{rs,r's'}\,U^*_{rk_1}V^*_{s'k_2}U_{r'k_3}V_{sk_4},
\label{eq:Xabc}
\end{align}
the vertices are
\begin{align}
 H^{40}_{k_1k_2k_3k_4}&=T_{1234}-T_{1324}+T_{1423}\nonumber\\
 &\quad+T_{2314}-T_{2413}+T_{3412},
 \label{eq:H40}\\
 H^{31}_{k_1k_2k_3k_4}&=\bigl[Y_{123;4}-Y_{213;4}+Y_{312;4}\bigr]\nonumber\\
 &\quad-\bigl[X_{123;4}-X_{132;4}+X_{231;4}\bigr],
 \label{eq:H31}\\
 H^{22}_{k_1k_2k_3k_4}&=X^a_{1234}+X^b_{1234}\nonumber\\
 &\quad-X^c_{1234}+X^c_{2134}+X^c_{1243}-X^c_{2143},
 \label{eq:H22}
\end{align}
where subscripts $1234$ abbreviate $k_1k_2k_3k_4$.  The vertex $H^{22}$ is Hermitian,
$H^{22}_{k_1k_2k_3k_4}=(H^{22}_{k_3k_4k_1k_2})^*$.
Expressions \eqref{eq:H40}--\eqref{eq:H22} were obtained by Wick's theorem. The derivation is given in Appendices~\ref{app:hfb} and \ref{app:vert}.

\begin{figure*}[t]
\centering
\includegraphics[width=0.8\textwidth]{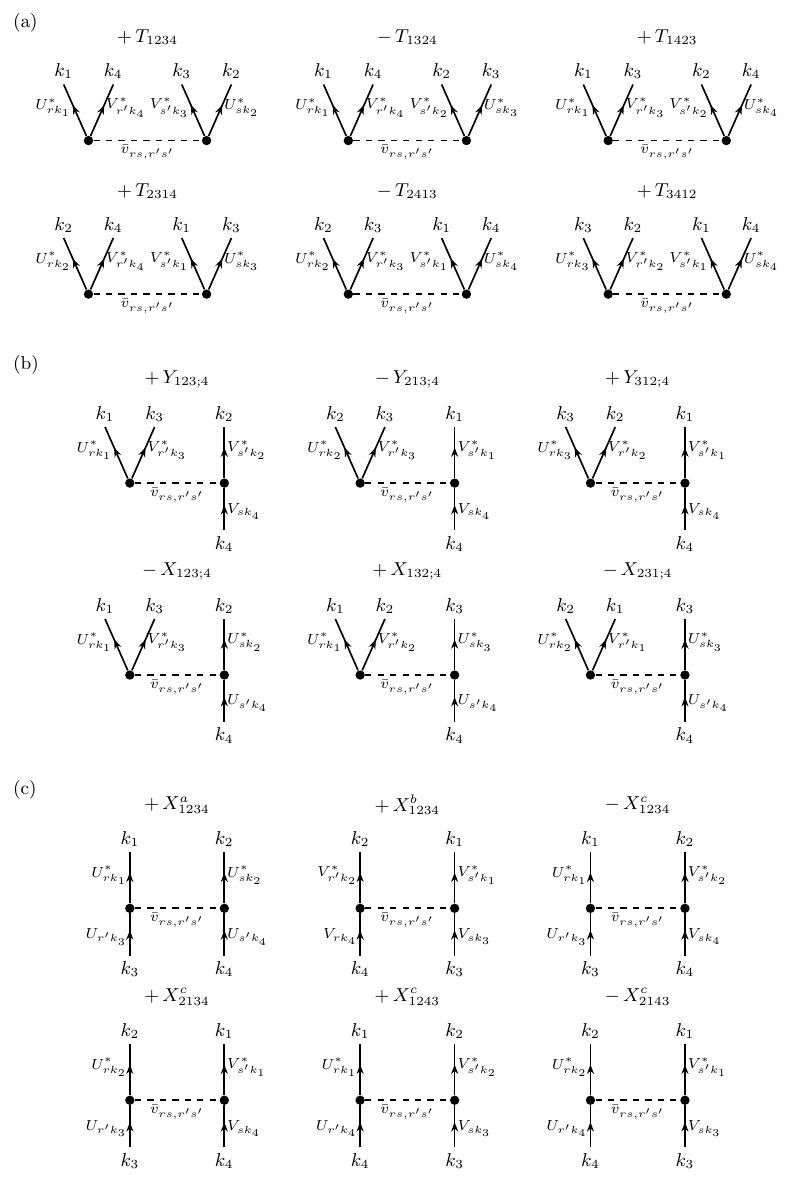}
\caption{Terms of the vertices (a) $H^{40}$, (b) $H^{31}$ and (c) $H^{22}$ (see Eqs.~\eqref{eq:H40}--\eqref{eq:H22}).
The dashed line is $\bar v_{rs,r's'}$ and time runs upward.  The sign of each diagram is that of the corresponding
term.}
\label{fig:vertexterms}
\end{figure*}

The terms of the vertices in Eqs. \eqref{eq:H40}--\eqref{eq:H22} are shown diagrammatically in Fig.~\ref{fig:vertexterms}.  A line leaving a point upward is a
created quasiparticle and a line entering from below an annihilated one.  By Eq.~\eqref{eq:invbog}, a quasiparticle
that originates from $a^\dagger$ carries the factor $U^*$ if it is created and $V$ if it is annihilated, and one that
originates from $a$ carries $V^*$ or $U$.  In $H^{22}$ the term $X^a$ contains only $U$ factors and $X^b$ only $V$
factors, while the terms $X^c$ carry $U$ at one point and $V$ at the other.

%=====================================================================
\section{SQRPA EQUATIONS}
\label{sec:SQRPA}
%=====================================================================

\subsection{Excitation operator and equation of motion}

We write the excitation operator as a superposition of two- and four-quasiparticle components,
\begin{equation}
 Q^\dagger_\nu=\sum_{a}\bigl(\mathcal{X}^\nu_a\,A^\dagger_a-\mathcal{Y}^\nu_a\,A_a\bigr),
\label{eq:Q}
\end{equation}
where the composite index $a$ runs over the ordered sets $a=(kk')$, $k<k'$ (two-quasiparticle space, 2qp)
and $a=(k_1k_2k_3k_4)$, $k_1<k_2<k_3<k_4$ (four-quasiparticle space, 4qp), and
\begin{align}
 A^\dagger_{(kk')}&=\ad k\ad{k'},\nonumber\\
 A^\dagger_{(k_1k_2k_3k_4)}&=\ad{k_1}\ad{k_2}\ad{k_3}\ad{k_4},
 \qquad A_a=(A^\dagger_a)^\dagger .
\end{align}
Because the index sets are ordered, the Pauli principle is built in.  Number-conserving
operators such as $\ad{}\ad{}\an{}\an{}$ annihilate $\vac$ and do not contribute in the QBA.

The equation of motion is
\begin{equation}
 \langle0|[\delta Q,[H,Q^\dagger_\nu]]|0\rangle=\omega_\nu\langle0|[\delta Q,Q^\dagger_\nu]|0\rangle,
 \label{eq:EOM}
\end{equation}
for $\delta Q\in\{A_a,A^\dagger_a\}$.  In the QBA, $|0\rangle\to\vac$, the
commutators on the right-hand side are exactly
$\bra[A_a,A^\dagger_b]\vac=\delta_{ab}$ and $\bra[A_a,A_b]\vac=0$.  With
\begin{equation}
\begin{aligned}
 \Amat_{ab}=&\bra A_a(H-E_0)A^\dagger_b\vac,
 \\
 \Bmat_{ab}=&\bra A_aA_bH\vac,
 \label{eq:ABdef}
\end{aligned} 
\end{equation}
Eq.~\eqref{eq:EOM} takes the standard form
\begin{equation}
 \begin{pmatrix}\Amat&\Bmat\\ \Bmat^*&\Amat^*\end{pmatrix}
 \begin{pmatrix}\mathcal X^\nu\\ \mathcal Y^\nu\end{pmatrix}=\omega_\nu
 \begin{pmatrix}\mathcal X^\nu\\-\mathcal Y^\nu\end{pmatrix},
 \label{eq:SQRPA}
\end{equation}
with normalization $\mathcal X^{\nu\dagger}\mathcal X^{\nu'}-\mathcal Y^{\nu\dagger}\mathcal Y^{\nu'}=\delta_{\nu\nu'}$ for
positive-norm modes.

\subsection{Structure of the matrices}

In the basis of the amplitudes $(\mathcal X_1,\mathcal X_2,\mathcal Y_1,\mathcal Y_2)$, where the indices $1$ and $2$ refer to the
2qp and the 4qp space, respectively, the matrix of Eq.~\eqref{eq:SQRPA} has the block structure
\begin{equation}
\Omega_{\text{SQRPA}}=\begin{pmatrix}
  \Amat_{11}&\Amat_{12}&\Bmat_{11}&0\\
  \Amat_{21}&\Amat_{22}&0&0\\
  \Bmat_{11}^*&0&\Amat_{11}^*&\Amat_{12}^*\\
  0&0&\Amat_{21}^*&\Amat_{22}^*
 \end{pmatrix}.
 \label{eq:Mfull}
\end{equation}
In the following, we discuss only which blocks vanish and present the explicit nonvanishing matrix elements in Sec.~\ref{nonzero_explicit}, whereas the full derivation of all blocks is relegated to Appendix~\ref{app:mat}.
Since $H\vac=E_0\vac+H^{40}\vac$, the matrix $\Bmat_{ab}$ requires removing the quasiparticles of $A_aA_b$
from a 4qp state, which is possible only if both $a$ and $b$ belong to the 2qp space. Therefore
\begin{equation}
 \Bmat_{12}=\Bmat_{21}=\Bmat_{22}=0 ,
 \label{eq:Bzero}
\end{equation}
and $\Bmat_{11}$ is symmetric. For $\Amat$, the quasiparticle numbers of the two configurations differ by $0$ within a sector and by $2$ between the
sectors.  A vertex $H^{mn}$, with $m$ creation and $n$ annihilation operators, changes the quasiparticle number by
$m-n$.  The vertices $H^{11}$ and $H^{22}$ therefore contribute only to $\Amat_{11}$ and $\Amat_{22}$, and $H^{31}$
and its conjugate only to $\Amat_{12}$ and $\Amat_{21}$.  The vertices $H^{40}$ and $H^{04}$ change the quasiparticle
number by $\pm4$ and connect the configurations only to the vacuum or to six-quasiparticle states, so they do not
contribute to $\Amat$.  They enter the SQRPA only through $\Bmat_{11}$.

\subsection{Explicit matrix elements}\label{nonzero_explicit}

Using the lemmas of Appendix~\ref{app:mat}, the blocks of $\Amat$ and $\Bmat$ can be written explicitly in terms of
the quasiparticle energies and the vertices $H^{22}$, $H^{31}$, and $H^{40}$.
Here $k<k'$ and $l<l'$ label the configurations of the 2qp space, and $\{k\}=(k_1<\dots<k_4)$ and
$\{l\}=(l_1<\dots<l_4)$ label those of the 4qp space. In $\Amat_{12}$ the indices $l_a<l_b<l_c$ are the three
elements of $\{l\}$ that remain after $l_d$ has been removed. One finds
\begin{align}
 \Amat_{11}(kk',ll')&=(E_k+E_{k'})\delta_{kl}\delta_{k'l'}+H^{22}_{kk'll'},\label{eq:A11}\\
 \Bmat_{11}(kk',ll')&=H^{40}_{kk'll'},\label{eq:B11}\\
 \Amat_{12}(kk',\{l\})&=\sum_{d=1}^{4}(-1)^d\Bigl[\bigl(H^{31}_{l_al_bl_c\,k}\bigr)^*\delta_{k'l_d}\nonumber\\
 &\qquad-\bigl(H^{31}_{l_al_bl_c\,k'}\bigr)^*\delta_{kl_d}\Bigr],\label{eq:A12}\\
 \Amat_{21}&=\Amat_{12}^\dagger,\nonumber\\
 \Amat_{22}(\{k\},\{l\})&=\Bigl(\sum_{i=1}^{4}E_{k_i}\Bigr)\delta_{\{k\}\{l\}}\nonumber\\
 &\quad+\sum_{i<j}\sum_{m<n}(-1)^{i+j+m+n}H^{22}_{k_ik_jl_ml_n}\nonumber\\
 &\qquad\times\delta_{\{k\}\setminus\{k_ik_j\},\{l\}\setminus\{l_ml_n\}}.\label{eq:A22}
\end{align}
In Eq.~\eqref{eq:A22}, $\delta_{\{k\}\{l\}}=\delta_{k_1l_1}\delta_{k_2l_2}\delta_{k_3l_3}\delta_{k_4l_4}$, since both
sets are ordered.
The symbol $\delta_{\{k\}\setminus\{k_ik_j\},\{l\}\setminus\{l_ml_n\}}$ equals one if the two quasiparticles
that remain in $\{k\}$ after removing $k_i$ and $k_j$ are the same as those that remain in $\{l\}$ after removing $l_m$ and $l_n$,
and zero otherwise. These remaining quasiparticles are spectators of the residual interaction $H^{22}$.
The block $\Amat_{22}$ thus describes the propagation of four quasiparticles, in which $H^{22}$ acts on a pair of them
while the other two are unaffected, and $\Amat_{11}$ and $\Bmat_{11}$ are the usual QRPA matrices.

\subsection{Limits and consistency}
\label{sec:limits}

Without pairing the Bogoliubov transformation reduces to the particle-hole transformation of Hartree--Fock (HF)
theory, and the SQRPA is expected to reduce to the SRPA.  In the basis that diagonalizes the HF field, with occupied
states $h$ ($\varepsilon_h<\mu$) and unoccupied states $p$ ($\varepsilon_p>\mu$), the only nonvanishing elements of the
transformation are $U_{rp}=\delta_{rp}$ and $V_{rh}=\delta_{rh}$, so that
\begin{equation}
 \alpha_p=a_p,\qquad \alpha_h=a^\dagger_h,\qquad \vac=|\mathrm{HF}\rangle ,
\end{equation}
and $E_p=\varepsilon_p-\mu$, $E_h=\mu-\varepsilon_h$.  Every factor $U$ in Eqs.~\eqref{eq:TXY} and \eqref{eq:Xabc}
selects a particle index and every factor $V$ a hole index, and only one term of each sum survives,
\begin{align}
 T_{k_1k_2k_3k_4}&=\vb_{k_1k_2,k_4k_3}, \quad k_1,k_2\in p,\ k_3,k_4\in h,\nonumber\\
 Y_{k_1k_2k_3;k_4}&=\vb_{k_1k_4,k_3k_2}, \quad k_1\in p,\ k_2,k_3,k_4\in h,\nonumber\\
 X_{k_1k_2k_3;k_4}&=\vb_{k_1k_2,k_3k_4}, \quad k_1,k_2,k_4\in p,\ k_3\in h,\nonumber\\
 X^a_{k_1k_2k_3k_4}&=\vb_{k_1k_2,k_3k_4}, \quad k_1,\dots,k_4\in p,\nonumber\\
 X^b_{k_1k_2k_3k_4}&=\vb_{k_4k_3,k_2k_1}, \quad k_1,\dots,k_4\in h,\nonumber\\
 X^c_{k_1k_2k_3k_4}&=\vb_{k_1k_4,k_3k_2}, \quad k_1,k_3\in p,\ k_2,k_4\in h,
 \label{eq:HFvert}
\end{align}
and they vanish otherwise.

A configuration with $n_p$ particle and $n_h$ hole quasiparticles changes the particle number by $n_p-n_h$.  Since $H$
conserves the particle number, $\Amat_{ab}$ vanishes unless $a$ and $b$ change it by the same amount, and
$\Bmat_{ab}$ vanishes unless the two changes add up to zero.  The configurations that conserve the particle number,
$1p1h$ in the two-quasiparticle space and $2p2h$ in the four-quasiparticle space, therefore decouple from the $2p$, $2h$, $3p1h$, $1p3h$, $4p$ and $4h$
configurations, which form separate particle-particle and hole-hole problems.  We identify the number-conserving
configurations with the operators
\begin{equation}
 O^\dagger_{ph}=a^\dagger_pa_h,\qquad O^\dagger_{p_1p_2h_1h_2}=a^\dagger_{p_1}a^\dagger_{p_2}a_{h_2}a_{h_1},
\end{equation}
with $p_1<p_2$ and $h_1<h_2$, which differ from $A^\dagger_a$ at most by a sign of each configuration and leave the eigenvalues unchanged.  With
Eqs.~\eqref{eq:H40}--\eqref{eq:H22} and \eqref{eq:HFvert}, Eqs.~\eqref{eq:A11}--\eqref{eq:A22} become
\begin{align}
 \Amat_{11}(ph,p'h')&=(\varepsilon_p-\varepsilon_h)\delta_{pp'}\delta_{hh'}+\vb_{ph',hp'},\label{eq:HFA11}\\
 \Bmat_{11}(ph,p'h')&=\vb_{pp',hh'},\label{eq:HFB11}\\
 \Amat_{12}(ph,p_1p_2h_1h_2)&=\delta_{hh_1}\vb_{ph_2,p_1p_2}-\delta_{hh_2}\vb_{ph_1,p_1p_2}\nonumber\\
 &\quad-\delta_{pp_1}\vb_{h_1h_2,hp_2}+\delta_{pp_2}\vb_{h_1h_2,hp_1},\label{eq:HFA12}
\end{align}
and
\begin{align}
 \Amat_{22}&(p_1p_2h_1h_2,p'_1p'_2h'_1h'_2)\nonumber\\
 ={}&(\varepsilon_{p_1}+\varepsilon_{p_2}-\varepsilon_{h_1}-\varepsilon_{h_2})\,\delta_{p_1p'_1}\delta_{p_2p'_2}\delta_{h_1h'_1}\delta_{h_2h'_2}\nonumber\\
 &+\vb_{p_1p_2,p'_1p'_2}\,\delta_{h_1h'_1}\delta_{h_2h'_2}
 +\vb_{h'_1h'_2,h_1h_2}\,\delta_{p_1p'_1}\delta_{p_2p'_2}\nonumber\\
 &+\sum_{i,j,k,l=1}^{2}(-1)^{i+j+k+l}\,\vb_{p_ih'_l,h_jp'_k}\,\delta_{p_{\bar\imath}p'_{\bar k}}\,\delta_{h_{\bar\jmath}h'_{\bar l}},
 \label{eq:HFA22}
\end{align}
where $\bar\imath$ denotes the index of the pair other than $i$. 
Because the configurations are ordered in Eq.~\eqref{eq:HFA22}, the antisymmetrizers of Ref.~\cite{Providencia} reduce to a single term in the
particle-particle and hole-hole contributions and give the explicit signed sum in the particle-hole contribution.
The blocks $\Bmat_{12}$, $\Bmat_{21}$ and $\Bmat_{22}$
remain zero by Eq.~\eqref{eq:Bzero}.  Equations~\eqref{eq:HFA11}--\eqref{eq:HFA22} are the matrices of the standard
SRPA with the full interaction in the $2p2h$ space, i.e. $\Amat_{ab}=\langle\mathrm{HF}|[O_a,[H,O^\dagger_b]]|\mathrm{HF}\rangle$ and
$\Bmat_{ab}=-\langle\mathrm{HF}|[O_a,[H,O_b]]|\mathrm{HF}\rangle$ (see also Refs.~\cite{Papakonstantinou2010}).  
The SQRPA thus reduces to the SRPA in the
number-conserving sector, while the particle-particle and hole-hole configurations of 2qp space give, together with
their 4qp space, a separate problem of the same structure for the neighbouring nuclei.

\subsection{Projection on good particle number}
\label{sec:projth}

The quasiparticle vacuum $\vac$ is not an eigenstate of $\hat N$, and neither are the states $Q^\dagger_\nu\vac$ built
on it.  The particle number is restored with the projector \cite{Sheikh2021}
\begin{equation}
 \hat P^N=\frac{1}{2\pi}\int_0^{2\pi}\!d\phi\,e^{i\phi(\hat N-N)} .
 \label{eq:PN}
\end{equation}
Because $[H,\hat N]=0$ and $\hat P^N$ is a Hermitian projector, all projected matrix elements between the vacuum and the
configurations $A^\dagger_a\vac$ follow from the norm and Hamiltonian kernels
$\langle a|e^{i\phi(\hat N-N)}|b\rangle$ and $\langle a|He^{i\phi(\hat N-N)}|b\rangle$,
\begin{equation}
 \mathsf N^P_{ab}=\langle a|\hat P^N|b\rangle,\qquad \mathsf H^P_{ab}=\langle a|H\hat P^N|b\rangle,
 \label{eq:NHP}
\end{equation}
where $a$ and $b$ run over the vacuum, $|0\rangle\equiv\vac$, and the two- and four-quasiparticle configurations. For the pairing
model of Sec.~\ref{sec:toy} the kernels are polynomials in $e^{2i\phi}$ and the integral in Eq.~\eqref{eq:PN} is
evaluated exactly with a finite number of gauge angles (see Appendix~\ref{app:proj}).

We use these matrices in two ways.  The first is the projection after variation of an (S)QRPA state.  In the
quasiboson approximation $A_a\vac=0$, so that $Q^\dagger_\nu\vac=\sum_{a}X^\nu_aA^\dagger_a\vac$ contains only the forward
amplitudes, and the backward amplitudes enter only through the values of $X^\nu_a$.  The projected state of the root is orthogonalized to the projected vacuum by adding a vacuum component,
\begin{equation}
\begin{aligned}
 |\nu^P\rangle=&\hat P^N\Bigl(c^\nu_0+\sum_aX^\nu_aA^\dagger_a\Bigr)\vac,\\
 c^\nu_0=&-\frac{1}{\mathsf N^P_{00}}\sum_b\mathsf N^P_{0b}X^\nu_b ,
 \label{eq:nuP}
\end{aligned} 
\end{equation}
so that $\bra\hat P^N|\nu^P\rangle=0$.  With $c^\nu_a=X^\nu_a$ for the configurations, the projected excitation
energy is
\begin{equation}
 \omega^P_\nu=\frac{\sum_{ij}c^{\nu*}_i\mathsf H^P_{ij}c^\nu_j}{\sum_{ij}c^{\nu*}_i\mathsf N^P_{ij}c^\nu_j}
 -\frac{\mathsf H^P_{00}}{\mathsf N^P_{00}},
 \label{eq:wP}
\end{equation}
where $i,j$ run over the vacuum and the configurations.  It measures the energy of the $N$-particle component of
the (S)QRPA state and is not an eigenvalue of the RPA equations. Without the orthogonalization the projected state would contain an admixture of the projected ground state and its
energy would be lowered accordingly.

The second is a projected configuration interaction (PCI) in the space $\mathcal V$ spanned by the vacuum and by the
two- and four-quasiparticle configurations (see Refs. \cite{Dietrich1964, Sheikh2021}.  The trial state $\hat P^N\bigl(c_0+\sum_{a}c_aA^\dagger_a\bigr)\vac$ leads to
the generalized eigenvalue problem (see also Ref. \cite{Hill1953})
\begin{equation}
 \sum_{b}\mathsf H^P_{ab}c_b=E\sum_{b}\mathsf N^P_{ab}c_b,\qquad a,b\in\{0\}\cup\mathcal C,
 \label{eq:PCI}
\end{equation}
where $\mathcal{C}$ denotes the set of 2qp and 4qp configurations and the excitation energies are $E_k-E_0$.  Without projection and without the vacuum, Eq.~\eqref{eq:PCI} reduces to
the eigenvalue problem of $\Amat$, which is the second Tamm--Dancoff approximation.  Including the vacuum adds its
couplings to the configurations. Without projection the coupling to the two-quasiparticle configurations vanishes because $H^{20}=0$, and the
coupling to the four-quasiparticle configurations is given by the vertex $H^{40}$, which enters the SQRPA through
$\Bmat_{11}$. In
Eq.~\eqref{eq:PCI} the ground-state correlations thus appear as a mixing of the vacuum with the configurations rather
than through backward amplitudes.  The projection removes the Goldstone direction exactly, since
$\hat P^N(\hat N-N)\vac=0$ makes it a null vector of $\mathsf N^P$, and Eq.~\eqref{eq:PCI} is solved on the range of
$\mathsf N^P$.  Because $\hat P^N\mathcal V$ is a subspace of the $N$-particle Hilbert space, the $k$th eigenvalue of
Eq.~\eqref{eq:PCI} is an upper bound to the $k$th exact eigenvalue.  The bound holds, to be precise, for total energies.  The
excitation energies have no such bound, because the ground state and the excited states are not described with the
same accuracy in a space truncated to two pair excitations.

Equation~\eqref{eq:PCI} is not a projected SQRPA.  The latter would require the projected kernels of the
de-excitation operators in the correlated ground state, which have not been formulated.  In this work the PCI serves as a reference that shows what the configuration space of the SQRPA can describe once
the particle number is conserved.

%=====================================================================
\section{ELIMINATION OF THE FOUR-QUASIPARTICLE SPACE AND SUBTRACTION}
\label{sec:red}
%=====================================================================

The coupling between the 2qp and 4qp spaces can be treated by eliminating the amplitudes of the 4qp space,
which reduces Eq.~(\ref{eq:SQRPA}) to an eigenvalue problem in the 2qp space alone.
Eliminating $\mathcal X_2$ and $\mathcal Y_2$ (details in Appendix~\ref{app:elim}) gives
\begin{equation}
\begin{aligned}
 \mathcal X_2=(\omega-\Amat_{22})^{-1}\Amat_{21}\mathcal X_1,\\
 \mathcal Y_2=-(\omega+\Amat_{22}^*)^{-1}\Amat_{21}^*\mathcal Y_1 ,
\end{aligned}
\end{equation}
and leaves an energy-dependent QRPA problem \cite{Tselyaev2013, Gambacurta2015},
\begin{equation}
 \begin{pmatrix}\Amat_{11}+\Sigma(\omega)&\Bmat_{11}\\
 \Bmat_{11}^*&\Amat_{11}^*+\tilde\Sigma(\omega)\end{pmatrix}
 \begin{pmatrix}\mathcal X_1\\ \mathcal Y_1\end{pmatrix}=\omega\begin{pmatrix}\mathcal X_1\\-\mathcal Y_1\end{pmatrix},
 \label{eq:eff}
\end{equation}
with
\begin{align}
 \Sigma(\omega)&=\Amat_{12}(\omega-\Amat_{22})^{-1}\Amat_{21},\nonumber\\
 \tilde\Sigma(\omega)&=-\Amat_{12}^*(\omega+\Amat_{22}^*)^{-1}\Amat_{21}^*.
\end{align}
The normalization condition becomes
\begin{equation}
\mathcal X_1^\dagger(1-\partial_\omega\Sigma)\mathcal X_1-\mathcal Y_1^\dagger(1+\partial_\omega\tilde\Sigma)\mathcal Y_1=1.
\end{equation}
The self-energies $\Sigma$ and $\tilde\Sigma$ describe the coupling of the two-quasiparticle configurations to virtual four-quasiparticle states.
Their derivatives with respect to $\omega$ determine the weight of the four-quasiparticle components of a mode, because from
the elimination one finds
$\mathcal X_2^\dagger\mathcal X_2=-\mathcal X_1^\dagger(\partial_\omega\Sigma)\mathcal X_1$ and
$\mathcal Y_2^\dagger\mathcal Y_2=\mathcal Y_1^\dagger(\partial_\omega\tilde\Sigma)\mathcal Y_1$.
The normalization of the full mode, which includes both spaces, then takes the form given above.
The evaluation of $\Sigma(\omega)$ requires the inversion of $\omega-\Amat_{22}$ in the 4qp space.
This is avoided in the diagonal approximation, in which only the diagonal elements of $\Amat_{22}$ are retained and
$\varepsilon_q\equiv\Amat_{22}(q,q)$ is the energy of the 4qp configuration $q$. The inverse is then trivial, and one has
\begin{align}
 \Sigma_{kk',ll'}(\omega)=\sum_q\frac{\Amat_{12}(kk',q)\,\Amat_{12}^*(ll',q)}{\omega-\varepsilon_q},\\
 \partial_\omega\Sigma_{kk',ll'}(\omega)=-\sum_q\frac{\Amat_{12}(kk',q)\,\Amat_{12}^*(ll',q)}{(\omega-\varepsilon_q)^2},
\end{align}
where $q$ runs over the 4qp configurations. Each term in $\Sigma$ is the contribution of one virtual 4qp state, and the
derivative is never positive for $k=l$ and $k'=l'$, so that the 4qp admixture adds to the norm of the mode.

\subsection{Subtraction}

For effective interactions (density functionals), second-order terms of the correlation energy are already
contained in the functional.  In the SRPA this leads to double counting, instabilities
(complex eigenvalues), and strong cutoff dependence, and the subtraction method removes them
by requiring that the static response coincide with that of the RPA \cite{Tselyaev2007,Gambacurta2015}.
The corresponding quasiparticle version of Eq.~\eqref{eq:eff} that we adopt replaces
\begin{equation}
 \Sigma(\omega)\to\Sigma(\omega)-\Sigma(0),\qquad
 \tilde\Sigma(\omega)\to\tilde\Sigma(\omega)-\tilde\Sigma(0).
 \label{eq:sub}
\end{equation}
Two properties follow directly from the definition.  First, the effective matrix at $\omega=0$
equals the QRPA matrix, so the static response of the subtracted SQRPA (SSQRPA) is the QRPA one.
Second, if $\xi_{\rm sp}=(\mathcal{X}_{\rm sp},\mathcal{Y}_{\rm sp})$ is a zero-energy solution of the QRPA (for instance the
Goldstone mode of broken particle-number symmetry), then $\omega=0$ with the same
2qp amplitudes is also a solution of the subtracted problem (Eqs. \eqref{eq:eff} and \eqref{eq:sub}).
Whether this mode coincides with the spurious mode of the full, unsubtracted SQRPA is
not addressed here.

Since $\Sigma(0)=-\Amat_{12}\Amat_{22}^{-1}\Amat_{21}$ does not depend on $\omega$,
the subtracted problem is identical to the ordinary SQRPA problem (Eq. \eqref{eq:SQRPA}, see also Ref. \cite{Vale2026} for detailed proof)
with the static replacement
\begin{equation}
 \Amat_{11}\;\to\;\Amat_{11}'=\Amat_{11}+\Amat_{12}\Amat_{22}^{-1}\Amat_{21}
 \label{eq:A11p}
\end{equation}
(and $\Amat_{11}^*\to\Amat_{11}'^{\,*}$ in the lower block), all other blocks being unchanged.
No energy-dependent equation has to be solved.  The counter-term in Eq.~\eqref{eq:A11p} equals $-\Sigma(0)$. It removes the static second-order shift of the
2qp energies caused by their coupling to the 4qp configurations.
Equation~\eqref{eq:sub} should be regarded as a proposal. Its validity in the superfluid case, in particular for the
$H^{31}$ vertex generated by a density-dependent functional, remains to be tested in realistic calculations.

\subsection{Size of the problem}

The number of 2qp configurations scales as $N^2/2$ and the number of
4qp configurations as $N^4/24$, where $N$ is the dimension of the
quasiparticle space. In realistic applications, cutoffs on the unperturbed 4qp energy
$\sum_iE_{k_i}$ and the use of angular-momentum-coupled bases would be unavoidable, as in the SRPA.
In the pairing model considered below no such truncation is needed.

%=====================================================================
\section{PAIRING MODEL WITH EQUALLY SPACED LEVELS}
\label{sec:toy}
%=====================================================================
To assess the quality of the approximation, 
we apply the SQRPA to a schematic pairing model whose exact solution can be obtained for small systems. The model
consists of $L$ equally spaced doubly degenerate levels filled with $N$ particles and interacting through a constant pairing force.
It is the simplest system in which the BCS vacuum breaks the particle-number symmetry, so that the QRPA has a spurious mode,
and in which the coupling to four-quasiparticle states can be tested against the exact result. We compare the QRPA, the SQRPA, and the subtracted SQRPA with the exact
solution for $L=N=6$ and $8$, and then examine how the differences scale with $L$ for $L=N=20$ and $28$.

\subsection{Model}

We take $L$ equidistant doubly degenerate levels, $\varepsilon_p=(p-1)d$ with $p=1,\dots,L$, each holding a
time-reversed pair of states $p$ and $\bar p$, with the pairing Hamiltonian
\begin{align}
 H=&\sum_p\varepsilon_p\,(n_p+n_{\bar p})-G\,P^\dagger P,\label{eq:Hpair}\\
 P^\dagger=&\sum_pa^\dagger_pa^\dagger_{\bar p},    
\end{align}
and $N=L$ particles (half filling).  The HFB vacuum is the BCS state obtained from the gap equation with the
Hartree--Fock term included, $\tilde\varepsilon_p=\varepsilon_p-Gv_p^2-\lambda$, and with $\lambda$ fixed by
$\langle\hat N\rangle=N$ (Eq.~\eqref{eq:gap} of Appendix~\ref{app:model}).  The vacuum is stationary for
$H'=H-\lambda\hat N$, as in Sec.~\ref{sec:HQP}.  
A paired solution exists above the critical coupling $G_c=0.424\,d$ ($L=6$) and $G_c=0.371\,d$ ($L=8$).  The values follow from Eq.~\eqref{eq:Gc} of Appendix~\ref{app:model} and agree with the numerical BCS
solution to within $2\times10^{-3}d$.  Below $G_c$ the solution is the normal-phase Hartree--Fock state and we do not use it.

Because $H$ does not break pairs, configurations with unpaired quasiparticles decouple from those in which every
quasiparticle has its time-reversed partner (Appendix~\ref{app:model}).  The two- and four-quasiparticle
configurations are therefore
\begin{equation}
 A^\dagger_p=\alpha^\dagger_p\alpha^\dagger_{\bar p},\qquad A^\dagger_{pq}=A^\dagger_pA^\dagger_q\quad(p<q),
 \label{ApApq}
\end{equation}
and the dimension of $\Amat$ is $L(L+1)/2$, that is $21$ for $L=6$ and $36$ for $L=8$.

\begin{figure*}[t]
\centering
\includegraphics[width=0.95\textwidth]{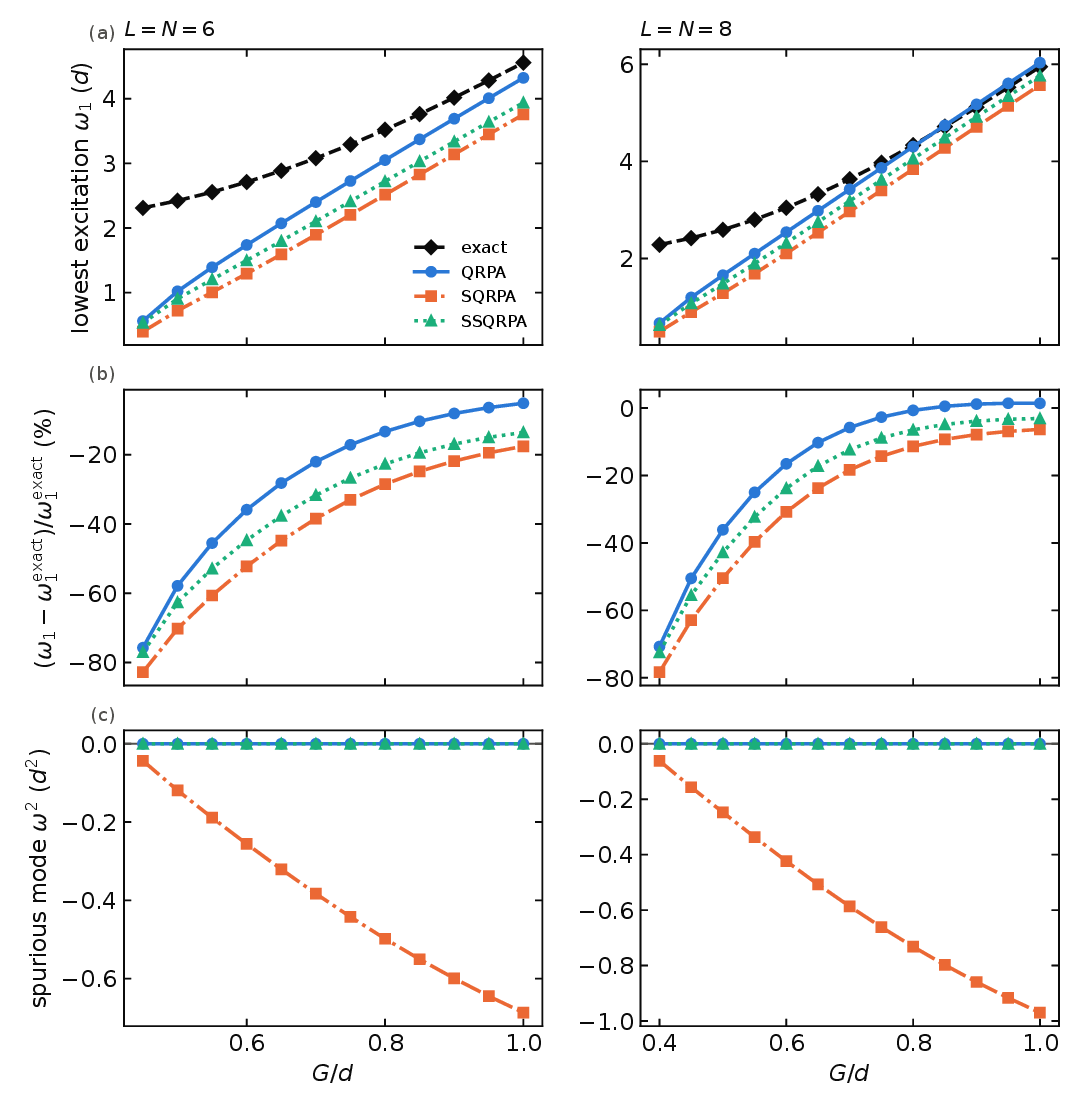}
\caption{Pairing model with $L=N=6$ (left) and $L=N=8$ (right).
(a)~Lowest excitation energy of the exact seniority-zero solution (diamonds), the QRPA, the SQRPA and
the subtracted SQRPA.  (b)~Relative deviation from the exact value.
(c)~Squared energy of the spurious mode associated with particle-number fluctuations, where
a negative value means an imaginary frequency.  For the QRPA and the subtracted SQRPA,
$|\omega^2|<10^{-12}\,d^2$ (the points lie on the zero line).}
\label{fig:toy}
\end{figure*}

For $L=6$ and $8$ at all couplings of Fig.~\ref{fig:toy} the blocks $\Bmat_{12}$, $\Bmat_{21}$ and $\Bmat_{22}$ computed
in the Fock space are at most $1.3\times10^{-15}$ in absolute value, as required by Eq.~\eqref{eq:Bzero}.  For all
sizes considered here the matrix $\Amat_{22}$ is positive definite, with its smallest eigenvalue at least $1.75\,d$
(reached for $L=6$ close to $G_c$), so that the elimination of Sec.~\ref{sec:red} is well defined.

The Hamiltonian~\eqref{eq:Hpair} conserves the particle number and does not break pairs.  The exact solution is
therefore obtained by diagonalizing $H$ in the space of the $\binom{L}{N/2}$ states in which the $N$ particles occupy
$N/2$ levels as time-reversed pairs (seniority zero, Appendix~\ref{app:exact}), and the exact excitation energy is $\omega_1^{\rm exact}=E_1-E_0$, where $E_0$ and $E_1$ are the energies of the
ground state and of the first excited state in this space. The configurations $A^\dagger_p\vac$ and
$A^\dagger_{pq}\vac$ also contain only pairs and lie in the same space.  States with broken pairs belong to other blocks
of $H$ and are not part of the comparison, although the lowest of them can lie below the first seniority-zero
excitation (Appendix~\ref{app:exact}).  We compare the exact excitation energy with the lowest excitation energy
$\omega_1$ of the QRPA (two-quasiparticle configurations only), of the SQRPA and of the SSQRPA, Eq.~\eqref{eq:A11p},
all obtained from Eq.~\eqref{eq:w2}.  The spurious mode of broken particle-number symmetry is identified as the
eigenvalue of $\omega^2$ closest to zero.  It is not counted in $\omega_1$ and is shown separately in
Fig.~\ref{fig:toy} and Table~\ref{tab:toy}.

\begin{table}[b]
\caption{Lowest excitation energy $\omega_1$ and squared frequency $\omega^2_{\rm sp}$ of the
spurious mode in the SQRPA (units $d$ and $d^2$).  For the QRPA and the SSQRPA,
$|\omega^2_{\rm sp}|<10^{-12}$.}
\label{tab:toy}
\begin{ruledtabular}
\begin{tabular}{ccccccc}
$L$ & $G/d$ & exact & QRPA & SQRPA & SSQRPA & $\omega^2_{\rm sp}$(SQRPA)\\
\hline
6 & 0.6 & 2.710 & 1.737 & 1.294 & 1.498 & $-0.256$\\
6 & 0.8 & 3.519 & 3.050 & 2.515 & 2.721 & $-0.498$\\
6 & 1.0 & 4.557 & 4.322 & 3.756 & 3.938 & $-0.687$\\
8 & 0.6 & 3.045 & 2.542 & 2.107 & 2.320 & $-0.424$\\
8 & 0.8 & 4.334 & 4.305 & 3.842 & 4.055 & $-0.732$\\
8 & 1.0 & 5.949 & 6.033 & 5.573 & 5.765 & $-0.970$\\
\end{tabular}
\end{ruledtabular}
\end{table}

\subsection{Results}

Figure~\ref{fig:toy} shows, as functions of $G$ for $L=6$ and $8$, the lowest excitation energy in the exact solution,
the QRPA, the SQRPA and the subtracted SQRPA, its relative deviation from the exact value, and the squared energy of
the spurious mode.  Table~\ref{tab:toy} lists the same quantities at $G/d=0.6$, $0.8$ and $1.0$.  We discuss the
spurious mode first and then the excitation energies and their interpretation.

(i) The QRPA has the zero-energy mode of broken particle-number symmetry
($\omega^2\approx10^{-13}d^2$).  In the unsubtracted SQRPA this mode becomes unstable, $\omega^2<0$, and
$|\omega^2|$ grows with $G$.  In the subtracted SQRPA it is again at zero, as required by the argument
of Sec.~\ref{sec:red}. A similar instability of the unsubtracted SQRPA, with complex eigenfrequencies, was found in realistic calculations
for oxygen isotopes~\cite{Yang2026}.

(ii) In the strongly paired region the QRPA is close to the exact result for
$L=8$ (within $2\%$ for $G\ge0.8\,d$) and $5$--$13\%$ below it for $L=6$ and $G\ge0.8\,d$.  The SQRPA lies lower
than the QRPA at every coupling, by about $0.4$--$0.5\,d$ for $L=8$ and $G\ge0.8\,d$, and is
therefore further from the exact result than the QRPA.  The subtracted SQRPA removes between
one third ($L=6$) and one half ($L=8$) of this shift but remains below the QRPA.  For weaker pairing, $G\lesssim0.6\,d$, all three
approximations fail badly (deviations of $-15\%$ to $-80\%$), as expected close to the transition from the normal phase,
where the mean-field vacuum is a poor description.

(iii) The downward shift is the behavior known for the SRPA built on an
uncorrelated reference state \cite{Tohyama2013}, i.e., the four-quasiparticle admixtures lower the two-quasiparticle
energies, but the ground state does not receive a matching correlation energy in the quasiboson
approximation.  Here the missing correlation energy is substantial,
$E_{\rm BCS}-E_{\rm exact}=1.5\,d$ ($L=6$) and $2.0\,d$ ($L=8$) at $G=d$.  We have not
tested this explanation by constructing a correlated ground state.

\subsection{Larger systems: \texorpdfstring{$L=N=20$ and $28$}{L=N=20 and 28}}
\label{sec:large}

The exact seniority-zero problem (Appendix~\ref{app:exact}) has $\binom{L}{L/2}$ states, about $1.8\times10^5$ for $L=20$ and $4\times10^7$ for
$L=28$. For $L=20$ its lowest eigenvalues are still obtained with a sparse eigensolver, and we use them below to test the
approximations. For $L=28$ no exact result is computed here. The Fock-space construction of $\Amat$ and $\Bmat$ used above is limited to $L\lesssim8$.
For larger $L$ we therefore construct the QRPA and SQRPA matrices in a smaller space.  In the space of pair
configurations each level $p$ is a two-state system, spanned by the quasiparticle vacuum of the level and by the state
with one quasiparticle pair, $A^\dagger_p=\alpha^\dagger_p\alpha^\dagger_{\bar p}$.  The pair operator
$a^\dagger_pa^\dagger_{\bar p}$ and the particle number $\hat N_p$ of the level are linear combinations of the raising,
lowering and number operators $\sigma^+_p$, $\sigma^-_p$ and $n_p$ of this two-state system (Eqs.~\eqref{eq:chainP} and
\eqref{eq:chainN} of Appendix~\ref{app:chain}).  The Hamiltonian $H'=H-\lambda\hat N$ is therefore a sum of products of
operators on a chain of $L$ two-state systems.
A matrix element between configurations with at most two excitations only involves
configurations with at most three, so $H'$ is built in a space of $1+L+\binom L2+\binom L3$ states ($3683$ for $L=28$), and
$\Amat$ and $\Bmat$ are read off from the definition \eqref{eq:ABdef} as before.  The sizes of $\Amat$ are $210\times210$
($L=20$) and $406\times406$ ($L=28$).  As a test, for $L=6$ and $8$ and $G/d=0.6,0.8,1.0$ this construction
reproduces the Fock-space matrices to $\max|\Delta\Amat|\le4\times10^{-13}$, $\max|\Delta\Bmat|\le2\times10^{-14}$, and the
frequencies of all three approximations.

Controls for $L=20$ and $28$ cover all couplings with a paired solution. They give $\max|H^{20}|\le1.3\times10^{-13}$. The matrices $\Bmat_{12}$, $\Bmat_{21}$ and $\Bmat_{22}$ vanish, while $\Amat_{22}$ is positive definite (smallest eigenvalue $\ge1.8\,d$). The spurious mode has $|\omega^2|\le6\times10^{-12}d^2$ in the QRPA and the subtracted SQRPA, and $-3.7\,d^2\le\omega^2<0$ in the unsubtracted SQRPA. The paired solution appears between $G=0.25\,d$ and $0.30\,d$ for $L=20$ and between $0.20\,d$ and $0.25\,d$ for $L=28$ (scan step $0.05\,d$).

\begin{figure*}[t]
\centering
\includegraphics[width=0.95\textwidth]{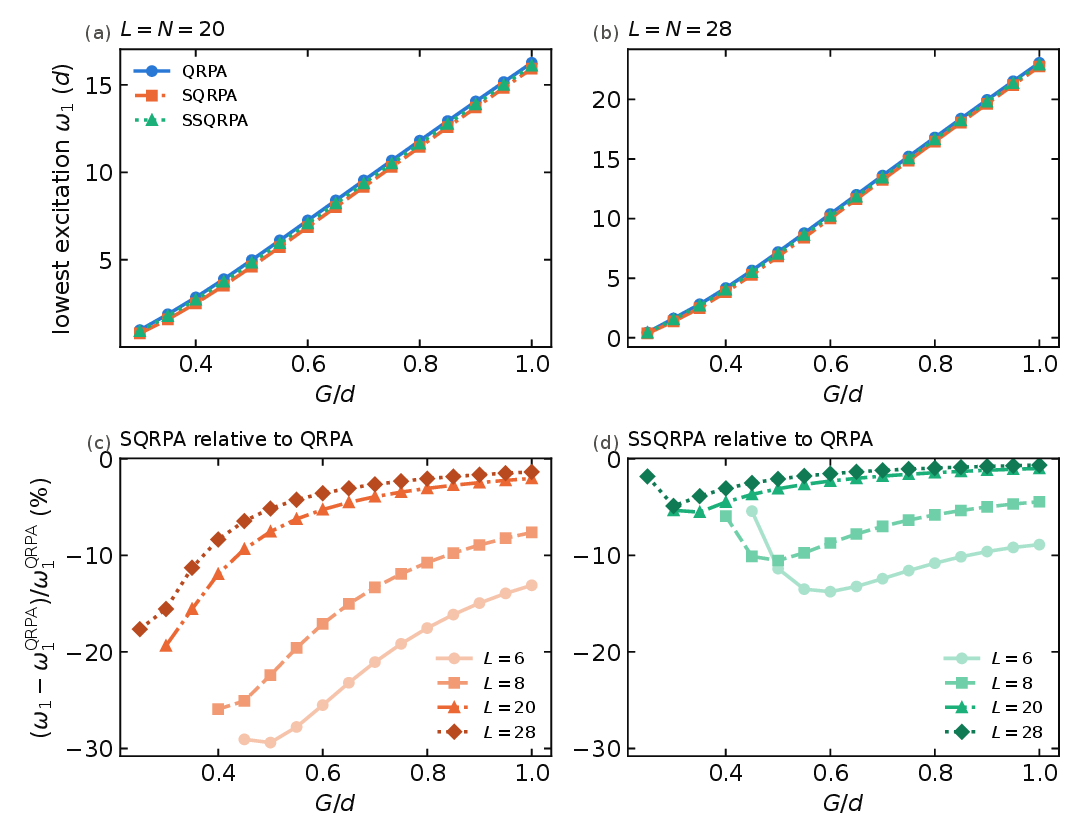}
\caption{Pairing model for larger systems.
(a) and (b) Lowest nonspurious excitation energy
$\omega_1$ in the QRPA, SQRPA, and SSQRPA for
$L=N=20$ and $28$, respectively.
(c) and (d) Relative shift of $\omega_1$ with
respect to the QRPA in the SQRPA and SSQRPA,
respectively, for $L=6$ and $8$ (Fock-space
calculation) and $L=20$ and $28$.}
\label{fig:large}
\end{figure*}

\begin{table*}[t]
\caption{Relative shift $(\omega_1-\omega_1^{\rm QRPA})/\omega_1^{\rm QRPA}$ (in \%) of the SQRPA and of the SSQRPA
with respect to the QRPA.}
\label{tab:large}
\begin{ruledtabular}
\begin{tabular}{c rr rr rr rr}
 & \multicolumn{2}{c}{$L=6$} & \multicolumn{2}{c}{$L=8$} & \multicolumn{2}{c}{$L=20$} & \multicolumn{2}{c}{$L=28$}\\
\cline{2-3}\cline{4-5}\cline{6-7}\cline{8-9}
$G/d$ & SQRPA & SSQRPA & SQRPA & SSQRPA & SQRPA & SSQRPA & SQRPA & SSQRPA \\
\hline
0.4 & --      & --      & $-25.9$ & $-5.9$  & $-11.9$ & $-4.5$ & $-8.4$ & $-3.1$\\
0.5 & $-29.4$ & $-11.4$ & $-22.4$ & $-10.5$ & $-7.5$  & $-3.1$ & $-5.2$ & $-2.1$\\
0.6 & $-25.5$ & $-13.8$ & $-17.1$ & $-8.7$  & $-5.3$  & $-2.3$ & $-3.6$ & $-1.5$\\
0.8 & $-17.5$ & $-10.8$ & $-10.7$ & $-5.8$  & $-3.1$  & $-1.4$ & $-2.1$ & $-1.0$\\
1.0 & $-13.1$ & $-8.9$  & $-7.6$  & $-4.4$  & $-2.0$  & $-1.0$ & $-1.4$ & $-0.6$\\
\end{tabular}
\end{ruledtabular}
\end{table*}

Figure~\ref{fig:large} and Table~\ref{tab:large} show that the qualitative picture does not change, i.e. (i) the SQRPA lies below
the QRPA, (ii) the unsubtracted version has an unstable spurious mode, and the (iii) subtraction restores it and reduces the shift
by a factor of two to three.  The size of the effect decreases quickly with $L$. At fixed $G$ the shift for $L=28$ is
about $0.7$ of that for $L=20$, close to the ratio $20/28$, i.e.\ it behaves roughly as $1/L$ between these two sizes (this
is an observation from two points, not a derived scaling law).  At $G=d$ the SQRPA shift is $-2.0\%$ for $L=20$ and $-1.4\%$
for $L=28$, and the subtracted one $-1.0\%$ and $-0.6\%$.  Since the shift tends to zero where the QRPA is expected to
become accurate, the results are consistent with the SQRPA correction being a finite-size effect in this model.
For $L=20$ the exact lowest nonspurious excitation energy is available (Table~\ref{tab:L20}). Near the onset of pairing, at $G=0.4\,d$,
all three approximations lie below the exact value and the QRPA is the closest, as for $L=6$ and $8$. For
$0.5\,d\le G\le0.9\,d$ the QRPA lies above the exact value by $0.7$--$10\%$. In this range the SSQRPA is closer to the
exact value than the QRPA, and so is the SQRPA up to $G=0.8\,d$. At $G=d$ the QRPA is again the closest.
The four-quasiparticle admixture can only lower $\omega_1$, so it improves on the QRPA only where the QRPA overestimates
the excitation energy by an amount comparable with the shift. For $L=8$ the QRPA lies above the exact value only for
$G\ge0.85\,d$, by at most $1.4\%$, which is much less than the shift. For $L=20$ the two are of similar size over most of
the paired region. For $L=28$ no exact result is available here.

\begin{table}[b]
\caption{Lowest nonspurious excitation energy $\omega_1$ (in units of $d$) for $L=N=20$. The exact value is the lowest
excitation of the seniority-zero spectrum (Appendix~\ref{app:exact}). The deviation
$(\omega_1-\omega_1^{\rm exact})/\omega_1^{\rm exact}$ (in \%) is given in parentheses.}
\label{tab:L20}
\begin{ruledtabular}
\begin{tabular}{ccccc}
$G/d$ & exact & QRPA & SQRPA & SSQRPA\\
\hline
0.4 & 3.009  & 2.851 ($-5.2$)  & 2.511 ($-16.5$) & 2.724 ($-9.5$)\\
0.5 & 4.524  & 4.987 ($+10.2$) & 4.611 ($+1.9$)  & 4.833 ($+6.8$)\\
0.6 & 6.682  & 7.250 ($+8.5$)  & 6.869 ($+2.8$)  & 7.084 ($+6.0$)\\
0.7 & 9.115  & 9.537 ($+4.6$)  & 9.163 ($+0.5$)  & 9.367 ($+2.8$)\\
0.8 & 11.565 & 11.809 ($+2.1$) & 11.447 ($-1.0$) & 11.639 ($+0.6$)\\
0.9 & 13.955 & 14.051 ($+0.7$) & 13.705 ($-1.8$) & 13.885 ($-0.5$)\\
1.0 & 16.286 & 16.261 ($-0.2$) & 15.931 ($-2.2$) & 16.101 ($-1.1$)\\
\end{tabular}
\end{ruledtabular}
\end{table}

\subsection{Particle-number projection}
\label{sec:proj}
 
The quasiparticle vacuum is a superposition of states with $N$, $N\pm2$, $N\pm4,\dots$ particles, and so are the
QRPA and SQRPA states built on it.  The exact levels, on the other hand, belong to the sector with fixed $N$.  To
separate the effect of number nonconservation from that of the coupling to the four-quasiparticle space, we use
the number-projected norm and Hamiltonian matrices of Appendix~\ref{app:proj}.  In the chain representation they
are exact and are obtained with $L+1$ gauge angles.
 
We consider two quantities.  The first is the projection after variation of the state $Q^\dagger_\nu\vac$ of each
root.  In the quasiboson approximation $A_p\vac=0$, so this state contains only the forward amplitudes $X_\nu$, and
the backward amplitudes enter only through the values of $X_\nu$.  The state is projected on $N$, orthogonalized to
the projected vacuum, and its energy $\omega^P_\nu$ is measured from the energy of the projected vacuum.  The weight
of the $N$-particle sector in these states lies between $0.1$ and $0.7$ for $L=8$ and $16$.  The second quantity is the
projected configuration mixing in the space $\mathcal V_2$ of the vacuum and of all configurations with one and two
quasiparticle pairs, which is the configuration space of the SQRPA.  It is variational and has no backward amplitudes.
Neither quantity is a projected SQRPA.
 
\begin{table}[t]
\caption{Lowest seniority-zero excitation energy (in units of $d$) for $L=N=8$ and $16$.  For each approximation the
unprojected frequency $\omega_1$ and the projected energy $\omega^P_1$ of the same root are given.  The last column is
the projected configuration mixing in $\mathcal V_2$ (Appendix~\ref{app:proj}).}
\label{tab:proj}
\begin{ruledtabular}
\begin{tabular}{llcccccccc}
 & & & \multicolumn{2}{c}{QRPA} & \multicolumn{2}{c}{SQRPA} & \multicolumn{2}{c}{SSQRPA} & \\
$L$ & $G/d$ & exact & $\omega_1$ & $\omega^P_1$ & $\omega_1$ & $\omega^P_1$ & $\omega_1$ & $\omega^P_1$ & $\mathcal V_2$\\
\hline
8  & 0.5 & 2.59 & 1.66 & 2.48 & 1.29 & 3.02 & 1.48 & 3.15 & 2.60\\
8  & 0.7 & 3.63 & 3.43 & 3.47 & 2.97 & 3.92 & 3.19 & 3.98 & 3.64\\
16 & 0.5 & 3.71 & 3.88 & 3.52 & 3.50 & 3.80 & 3.73 & 3.83 & 3.76\\
16 & 0.7 & 7.04 & 7.50 & 7.23 & 7.11 & 7.14 & 7.32 & 7.14 & 7.29\\
\end{tabular}
\end{ruledtabular}
\end{table}
 
\begin{figure*}[t]
\centering
\includegraphics[width=0.8\textwidth]{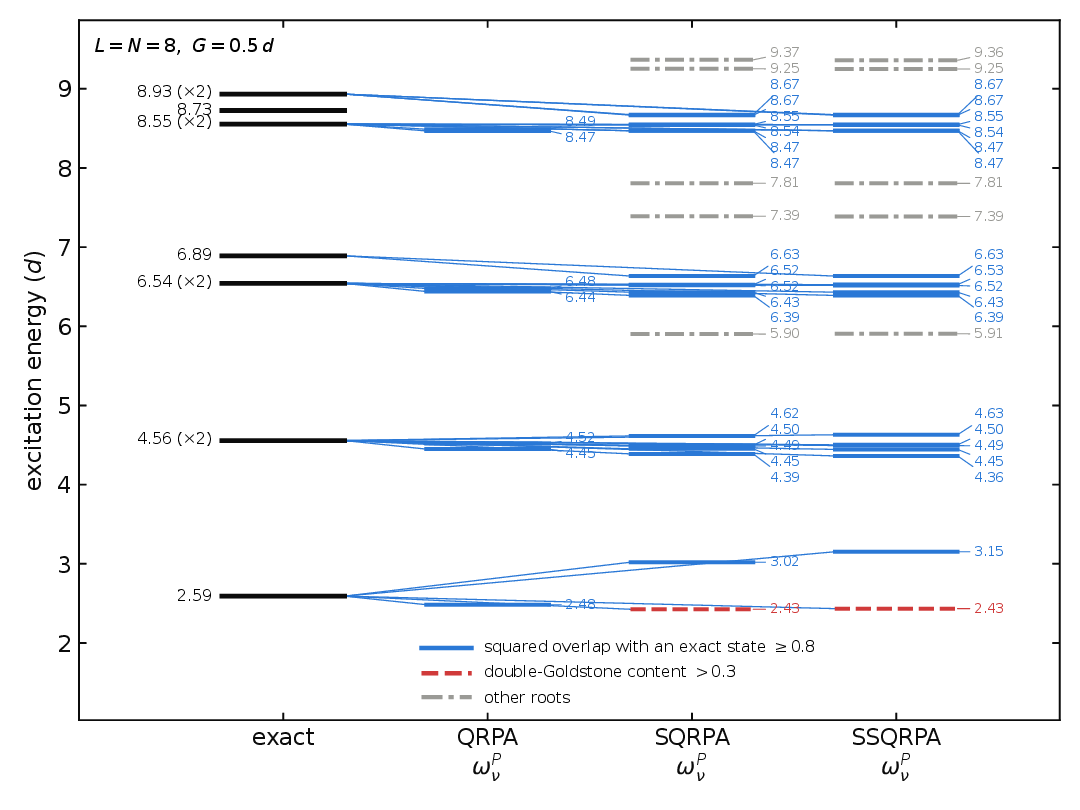}
\caption{Excitation energies up to $9.5\,d$ for
$L=N=8$ and $G=0.5\,d$. The leftmost column
shows the exact seniority-zero spectrum, with
degeneracies indicated in parentheses. The
remaining columns show the projected excitation
energies $\omega_\nu^P$ associated with the QRPA,
SQRPA, and SSQRPA roots. Solid (blue) lines
identify projected states with a squared overlap
of at least $0.8$ with an exact eigenstate.
Thin lines connect these states to their
corresponding exact levels. Dashed (red) lines
indicate roots with a double-Goldstone content
$|\langle g_2|\nu\rangle|^2>0.3$, where
$|g_2\rangle$ is the normalized four-quasiparticle
component of $(\hat N-N)^2\vac$ and
$|\nu\rangle=\sum_a X_a^\nu A_a^\dagger\vac$.
The remaining roots are shown by dash-dotted
(grey) lines.}
\label{fig:proj}
\end{figure*}
 
Table~\ref{tab:proj} lists the lowest excitation energy.  For $L=8$ the projection raises the lowest QRPA state from
$1.66\,d$ to $2.48\,d$, close to the exact $2.59\,d$.  The lowest SQRPA and SSQRPA roots, which before
projection lie further below the exact level than the QRPA, lie above it after projection, at $3.02\,d$ and $3.15\,d$.
The large downward shift of the unprojected SQRPA frequency at $L=8$ is therefore not present in the projected
energy of the same state, which suggests that it is associated with the components of other particle numbers.  For
$L=16$ the differences between the three approximations are smaller, and at $G/d=0.7$ the projected SQRPA state is the
closest to the exact level ($7.14\,d$ against $7.04\,d$).
 
Figure~\ref{fig:proj} shows the higher states for $L=8$ and $G/d=0.5$.  After projection most of the additional SQRPA
roots fall close to the exact levels at $4.56\,d$ and $6.54\,d$, each of which is matched by several roots.  The higher
density of SQRPA roots below $8\,d$ thus comes largely from roots that represent the same $N$-particle state.  The root
with the largest double-Goldstone content behaves differently for the two sizes.  For $L=8$ its projection has an
overlap of $0.94$ with the lowest exact excited state, while for $L=16$ ($G/d=0.5$) its weight in the $N$-particle sector
is $0.20$ and its projection overlaps the projected vacuum with $0.66$.
 
The projected configuration mixing in $\mathcal V_2$ reproduces the exact excitation energies to within $0.05\,d$ for
$L=8$ and $0.25\,d$ for $L=16$, and the ground-state energy to within $10^{-2}d$.  The space of configurations with up to
two quasiparticle pairs is therefore sufficient to describe the low-lying $N$-particle states of this model, and the
deviations of the SQRPA are not due to the size of the space.  Possible causes are the quasiboson approximation, the
uncorrelated reference state and the nonconservation of the particle number.  We have not separated these contributions.
A projected SQRPA, in which the backward amplitudes and the correlated ground state are treated with projected kernels,
has not been formulated and is left for future work.

\subsection{Limitations}

The model has a bare interaction, not an effective one, so the arguments for the subtraction method
(double counting with a density functional) do not apply directly, i,e, its effect here is that of a static
counter-term restoring the zero mode.  The comparison with the exact solution identifies QRPA modes
with excitations at fixed particle number, which is an approximation for small systems.
The exact reference is restricted to seniority zero. The lowest state of the full spectrum at fixed $N$ may have broken pairs,
which we have not computed.
The model has no particle--hole channel and its excitations are pair vibrations of a single kind. A model with particle--hole correlations, and nuclei with a realistic interaction, are needed before
conclusions about the usefulness of the SQRPA can be drawn.

\section{SUMMARY AND OUTLOOK}
\label{sec:sum}

We have formulated the second QRPA for a general Bogoliubov vacuum.  Starting from the equation of motion in the
quasiboson approximation, we wrote the complete quasiparticle vertices $H^{40}$, $H^{31}$ and $H^{22}$ and the
matrices $\Amat$ and $\Bmat$ explicitly for arbitrary transformation matrices $U$ and $V$.  The structure of the
problem follows from the change of the quasiparticle number.  The matrix $\Bmat$ is nonzero only between
two-quasiparticle configurations, the coupling to the four-quasiparticle space is determined by $H^{31}$ alone, and
the elimination of that space gives an energy-dependent QRPA to which a subtraction can be applied.  Without pairing
the equations reduce to the standard SRPA.  All formulas were verified to machine precision in an independent
Fock-space calculation.  The equations are therefore a general starting point for second-order quasiparticle
calculations in open-shell nuclei, not a model-specific construction.

The picket-fence pairing model served as a first test with an exact solution.  In this model the SQRPA finds the
right states but lowers their energies below the QRPA and the exact result, and the subtraction reduces this shift
without removing it.  The comparison with the projected calculations provides insight into the origin of these deviations. The space of two- and four-quasiparticle configurations is sufficient to describe the low-lying states of the pairing model with good accuracy, as demonstrated by the projected configuration interaction. This suggests that the SQRPA deviations arise primarily from the approximations used to treat this space, rather than from its size. Possible sources include the quasiboson approximation, the uncorrelated reference state, and the breaking of particle-number symmetry. Their individual contributions have not been disentangled. For
$L=20$ and $28$ the shifts relative to the QRPA decrease roughly as $1/L$.  Whether the shift improves on the QRPA
depends on the sign of the QRPA error.  Since the second-order coupling can only lower the lowest root, the SQRPA
worsens the QRPA result where it is already close to or below the exact value, as for $L=6$ and $8$, and improves on it
where the QRPA overestimates the excitation energy, as for $L=20$ at intermediate coupling.  The latter situation is
consistent with Ref.~\cite{Yang2026}, where the four-quasiparticle configurations bring the low-lying dipole states of
the oxygen isotopes closer to experiment.  A schematic model with a pure pairing force is a severe test for any method
built on a number-breaking vacuum, and the results should be read as a benchmark of the formalism rather than as a
verdict on its use in realistic nuclei.

This work is a first step, and several directions follow directly from it.  The vertices can be evaluated with
effective interactions and energy density functionals, including relativistic ones, which requires the choice of
configuration cutoffs and a consistent subtraction in the superfluid case.  The quasiboson approximation can be
relaxed by a correlated ground state or by renormalized and self-consistent extensions, which would supply the
ground-state correlations that the present scheme lacks.  The particle number can be restored within the equations of
motion rather than after variation, which would lead to a projected SQRPA.  The pairing model, with its exact
solution and the chain representation for large $L$, provides a controlled test for each of these extensions.
%=====================================================================
\appendix
\section{BOGOLIUBOV TRANSFORMATION, VACUUM AND CONTRACTIONS}
\label{app:hfb}
%=====================================================================
This appendix collects the properties of the Bogoliubov--Valatin transformation \eqref{eq:bog} that are used in the main text.
We first show that the canonical conditions on the matrices $U$ and $V$ are equivalent to the fermionic anticommutation
relations of the quasiparticle operators, and derive the inverse transformation. We then obtain the normal and pairing
densities $\rho$ and $\kappa$ of the vacuum $\vac$ and discuss the freedom in the choice of the quasiparticle basis.

\subsection{Matrix form and canonical conditions}

We first express the Bogoliubov--Valatin transformation (\ref{eq:bog}) in matrix form and determine which conditions the matrices $U$ and $V$ must satisfy so that the operators $\alpha_k$ and $\alpha^\dagger_k$ obey the fermionic anticommutation relations.
Collecting Eq.~(\ref{eq:bog}) for all $k$,
\begin{equation}
 \begin{pmatrix}\alpha\\ \alpha^\dagger\end{pmatrix}=\mathcal W^\dagger
 \begin{pmatrix}a\\ a^\dagger\end{pmatrix},\qquad
 \mathcal W=\begin{pmatrix}U&V^*\\ V&U^*\end{pmatrix}.
 \label{eq:Wmat}
\end{equation}
The first block row of $\mathcal W^\dagger=\bigl(\begin{smallmatrix}U^\dagger&V^\dagger\\ V^T&U^T\end{smallmatrix}\bigr)$
gives $\alpha_k=\sum_r(U^*_{rk}a_r+V^*_{rk}a^\dagger_r)$, and the second row is its adjoint.  With
$\{a_r,a^\dagger_{r'}\}=\delta_{rr'}$ and $\{a_r,a_{r'}\}=0$ one finds
\begin{align}
 \{\alpha_k,\alpha^\dagger_l\}&=\sum_r\bigl(U^*_{rk}U_{rl}+V^*_{rk}V_{rl}\bigr)=(U^\dagger U+V^\dagger V)_{kl},\nonumber\\
 \{\alpha_k,\alpha_l\}&=\sum_r\bigl(U^*_{rk}V^*_{rl}+V^*_{rk}U^*_{rl}\bigr)=(U^\dagger V^*+V^\dagger U^*)_{kl}.
\end{align}
The anticommutation relations for the $\alpha_k$ hold if and only if $\mathcal W^\dagger\mathcal W=1$, i.e.\
\begin{equation}
 U^\dagger U+V^\dagger V=1,\qquad U^\dagger V^*+V^\dagger U^*=0 .
 \label{eq:canon1}
\end{equation}
Since $\mathcal W$ is a square matrix, this implies $\mathcal W\mathcal W^\dagger=1$, which gives
\begin{equation}
 UU^\dagger+V^*V^T=1,\qquad UV^\dagger+V^*U^T=0 ,
 \label{eq:canon2}
\end{equation}
and the inverse transformation $(a,a^\dagger)^T=\mathcal W(\alpha,\alpha^\dagger)^T$ is given by Eq.~(\ref{eq:invbog}).

\subsection{Densities and contractions}

Below $\langle X\rangle\equiv\bra X\vac$.  In any bilinear of the operators (\ref{eq:invbog}) only
$\langle\alpha_k\alpha^\dagger_l\rangle=\delta_{kl}$ survives, because $\alpha_k\vac=0$ and $\bra\ad k=0$.  Hence
\begin{align}
 \rho_{rr'}&\equiv\langle a^\dagger_{r'}a_r\rangle
 =\sum_{kl}\bigl\langle(U^*_{r'k}\ad k+V_{r'k}\an k)(U_{rl}\an l+V^*_{rl}\ad l)\bigr\rangle\nonumber\\
 &=\sum_kV_{r'k}V^*_{rk}=(V^*V^T)_{rr'},\label{eq:rho}\\
 \kappa_{rr'}&\equiv\langle a_{r'}a_r\rangle
 =\sum_{kl}\bigl\langle(U_{r'k}\an k+V^*_{r'k}\ad k)(U_{rl}\an l+V^*_{rl}\ad l)\bigr\rangle\nonumber\\
 &=\sum_kU_{r'k}V^*_{rk}=(V^*U^T)_{rr'} .\label{eq:kappa}
\end{align}
So $\rho$ is Hermitian, $\kappa$ is antisymmetric, and the third contraction is
$\langle a^\dagger_ra^\dagger_{r'}\rangle=\langle a_{r'}a_r\rangle^*=\kappa^*_{rr'}$.
These three are the only contractions needed below, because every product that occurs has the
creation operators $a^\dagger$ to the left of the annihilation operators $a$.

%=====================================================================
\section{NORMAL-ORDERED HAMILTONIAN AND VERTICES}
\label{app:vert}
%=====================================================================

We use $H$ of Eq.~(\ref{eq:H}), with $\tilde t$ Hermitian and $\vb_{rs,r's'}$ antisymmetric in $(rs)$ and in $(r's')$ and obeying
$\vb_{rs,r's'}=\vb^*_{r's',rs}$ (see Refs. \cite{RingSchuck1980, Suhonen2007}).

\subsection{Wick's theorem relative to the quasiparticle vacuum}

The symbol $:\!X\!:$ denotes the normal-ordered product of $X$ with respect to $\alpha,\alpha^\dagger$. It is obtained by placing all $\alpha^\dagger$ to the left of all $\alpha$ and multiplying by the sign of the permutation that produces this order, with no anticommutator terms.  Then $\bra:\!X\!:\vac=0$ and
\begin{equation}
 a_1a_2\cdots a_n=\;:\!a_1\cdots a_n\!:\;+\sum_{\rm single}\;+\sum_{\rm double}+\cdots,
\end{equation}
where a term with contraction $\langle a_ia_j\rangle$ ($i<j$) carries the sign $(-1)^{j-i-1}$ needed to bring $a_j$ next to $a_i$,
and the remaining operators are normal ordered.  Since every $a_r$, $a^\dagger_r$ is a sum of one $\alpha$-term and one $\alpha^\dagger$-term,
Eq.~(\ref{eq:invbog}), the normal-ordered product of $a$'s is obtained by replacing each $a$ by either of its terms
(a ``channel'') and reordering with the permutation sign.

\subsection{Contractions and mean fields}

Wick's theorem is used to bring $H$ to normal-ordered form with respect to the quasiparticle vacuum.
For the one-body term one has
$a^\dagger_ra_{r'}=\rho_{r'r}+:\!a^\dagger_ra_{r'}\!:$ .
The product $a^\dagger_ra^\dagger_sa_{s'}a_{r'}$ in the two-body term has six single contractions (Table~\ref{tab:contr}).

\begin{table}[b]
\caption{Single contractions of $a^\dagger_ra^\dagger_sa_{s'}a_{r'}$ (operators labeled 1--4 from the left).
Sign $=(-1)^{j-i-1}$ for contraction of operators $i<j$.}
\label{tab:contr}
\begin{ruledtabular}
\begin{tabular}{cccc}
Pair & Value & Sign & Remainder\\
\hline
(1,4) & $\rho_{r'r}$ & $+$ & $:\!a^\dagger_sa_{s'}\!:$\\
(2,3) & $\rho_{s's}$ & $+$ & $:\!a^\dagger_ra_{r'}\!:$\\
(1,3) & $\rho_{s'r}$ & $-$ & $:\!a^\dagger_sa_{r'}\!:$\\
(2,4) & $\rho_{r's}$ & $-$ & $:\!a^\dagger_ra_{s'}\!:$\\
(3,4) & $\kappa_{r's'}$ & $+$ & $:\!a^\dagger_ra^\dagger_s\!:$\\
(1,2) & $\kappa^*_{rs}$ & $+$ & $:\!a_{s'}a_{r'}\!:$\\
\end{tabular}
\end{ruledtabular}
\end{table}

Multiplying by $\tfrac14\vb_{rs,r's'}$ and using the antisymmetry of $\vb$ in $(rs)$ and $(r's')$, the four $\rho$-terms are equal after
relabeling ($r\leftrightarrow s$, $r'\leftrightarrow s'$) and add up to a single term with coefficient~1. We define the Hartree--Fock field $\Gamma$ and the pairing field $\Delta$,
\begin{align}
\Gamma_{rr'}&=\sum_{ss'}\vb_{sr,s'r'}\,\rho_{s's}=\sum_{ss'}\vb_{rs,r's'}\,\rho_{s's},\nonumber\\
 \Delta_{rs}&=\tfrac12\sum_{r's'}\vb_{rs,r's'}\,\kappa_{r's'}.
 \label{eq:GamDel}
\end{align}
Then the two $\kappa$-terms give $\tfrac12\sum\Delta_{rs}:\!a^\dagger_ra^\dagger_s\!:$ and its Hermitian conjugate (the conjugate follows from
$\vb_{rs,r's'}=\vb^*_{r's',rs}$).  The double contractions are $(1,4)(2,3)$, $(1,3)(2,4)$ and $(1,2)(3,4)$, The first two are equal and
the third is $\tfrac14\sum\vb\,\kappa^*_{rs}\kappa_{r's'}$.  Altogether
\begin{align}
 H&=E_0+\sum_{rr'}h_{rr'}:\!a^\dagger_ra_{r'}\!:\;+\tfrac12\sum_{rs}\bigl(\Delta_{rs}:\!a^\dagger_ra^\dagger_s\!:+{\rm h.c.}\bigr)\nonumber\\
 &\quad+\tfrac14\sum_{rsr's'}\vb_{rs,r's'}:\!a^\dagger_ra^\dagger_sa_{s'}a_{r'}\!:\;,\label{eq:Hwick}\\
 h&=\tilde t+\Gamma,\nonumber\\
 E_0&=\sum_{rr'}\bigl(\tilde t_{rr'}+\tfrac12\Gamma_{rr'}\bigr)\rho_{r'r}+\tfrac12\sum_{rs}\Delta_{rs}\kappa^*_{rs}.\label{eq:E0f}
\end{align}
Equivalently, the last term of $E_0$ is $-\tfrac12{\rm Tr}\,\Delta\kappa^*$.

\subsection{The two-operator part in the quasiparticle basis}

Inserting the channels of Eq.~\eqref{eq:invbog} into $:\!a^\dagger_ra_{r'}\!:$ gives four terms.  In the term
$V_{rk}\an k\,V^*_{r'l}\ad l$ the annihilation operator stands to the left, and normal ordering brings it to
$-V_{rk}V^*_{r'l}\ad l\an k$.  Collecting the terms, one obtains
\begin{align}
 \sum_{rr'}h_{rr'}:\!a^\dagger_ra_{r'}\!:&=\sum_{kl}\bigl(U^\dagger hU-V^\dagger h^TV\bigr)_{kl}\ad k\an l\nonumber\\
 &\quad+\sum_{kl}(U^\dagger hV^*)_{kl}\ad k\ad l+(\an{}\an{}\ \text{terms}).
\end{align}
For the pairing field, $\tfrac12\sum_{rs}\Delta_{rs}:\!a^\dagger_ra^\dagger_s\!:$ gives, from the channels in which both operators are replaced by creation operators,
\begin{equation}
 \tfrac12\sum_{kl}\bigl(U^\dagger\Delta U^*\bigr)_{kl}\,\ad k\ad l ,
\end{equation}
and, from the mixed channels,
\begin{align}
 &\tfrac12\sum_{kl}\Bigl[\bigl(U^\dagger\Delta V\bigr)_{kl}-\bigl(V^T\Delta U^*\bigr)_{lk}\Bigr]\ad k\an l\nonumber\\
 &\qquad=\sum_{kl}\bigl(U^\dagger\Delta V\bigr)_{kl}\,\ad k\an l ,
\end{align}
where we used $\Delta^T=-\Delta$. The Hermitian conjugate of the pairing field contributes
\begin{align}
 &-\sum_{kl}\bigl(V^\dagger\Delta^*U\bigr)_{kl}\,\ad k\an l ,\\
 &-\tfrac12\sum_{kl}\bigl(V^\dagger\Delta^*V^*\bigr)_{kl}\,\ad k\ad l .
\end{align}
Collecting,
\begin{align}
 e&=U^\dagger hU-V^\dagger h^TV+U^\dagger\Delta V-V^\dagger\Delta^*U,\label{eq:ef}\\
 H^{20}&=U^\dagger hV^*-(U^\dagger hV^*)^T+U^\dagger\Delta U^*-V^\dagger\Delta^*V^*,\label{eq:H20f}
\end{align}
with $H^{11}=\sum e_{kl}\ad k\an l$ and $H^{20}=\tfrac12\sum H^{20}_{kl}\ad k\ad l$ (the antisymmetrization in the first
line of (\ref{eq:H20f}) comes from $\ad k\ad l=-\ad l\ad k$).  The condition $H^{20}=0$ is the HFB equation.

\subsection{The four-operator part}

In $\tfrac14\sum\vb_{rs,r's'}:\!a^\dagger_ra^\dagger_sa_{s'}a_{r'}\!:$ each of the four operators is replaced by one of its two channels:
$a^\dagger_r\to U^*_{rk}\ad k$ or $V_{rk}\an k$, and $a_r\to U_{rk}\an k$ or $V^*_{rk}\ad k$.  This gives $2^4=16$ terms, i.e. one with four $\alpha^\dagger$, four with
three $\alpha^\dagger$ and one $\alpha$, six with two of each, four with one $\alpha^\dagger$ and three $\alpha$, and one with four $\alpha$.  The last two groups are the
Hermitian conjugates of the first two.  In every term the $\alpha^\dagger$ indices are named $k_1,k_2,(k_3)$ in the order in which the operators appear
in the product and the $\alpha$ index/indices $k_4,(k_3)$ likewise, the first appearing one getting the larger label. The term is then rewritten with the
operators in the standard order $\ad{k_1}\ad{k_2}\ad{k_3}\an{k_4}$ or $\ad{k_1}\ad{k_2}\an{k_4}\an{k_3}$.  Relabeling of dummy summation
indices ($r\leftrightarrow s$ or $r'\leftrightarrow s'$) brings the coefficients to the tensors (\ref{eq:TXY}) and (\ref{eq:Xabc}) at the price of a sign from the antisymmetry of $\vb$.

The channel in which all four operators give creation operators is already in standard order and contributes
$\tfrac14\sum_{k_1k_2k_3k_4}T_{k_1k_2k_3k_4}\,\ad{k_1}\ad{k_2}\ad{k_3}\ad{k_4}$, with $T$ defined in Eq.~(\ref{eq:TXY}).

\begin{table}[tb]
\caption{Channels of $\tfrac14\sum\vb_{rs,r's'}\,:\!a^\dagger_ra^\dagger_sa_{s'}a_{r'}\!:$ with three quasiparticle
creation operators and one annihilation operator.  For each channel the table gives the sign of the reordering to
$\ad{k_1}\ad{k_2}\ad{k_3}\an{k_4}$, the tensor obtained after relabeling, and their product.}
\label{tab:31}
\begin{ruledtabular}
\begin{tabular}{lccc}
Operator giving $\alpha$ & Sign & Tensor & Net\\
\hline
$a^\dagger_r$ & $-$ & $-Y_{k_1k_2k_3;k_4}$ ($r\leftrightarrow s$) & $+Y$\\
$a^\dagger_s$ & $+$ & $+Y_{k_1k_2k_3;k_4}$ & $+Y$\\
$a_{s'}$ & $-$ & $+X_{k_1k_2k_3;k_4}$ & $-X$\\
$a_{r'}$ & $+$ & $-X_{k_1k_2k_3;k_4}$ ($r'\leftrightarrow s'$) & $-X$\\
\end{tabular}
\end{ruledtabular}
\end{table}

The channels in which one of the four operators gives an annihilation operator and the other three give creation operators are listed in Table~\ref{tab:31}.
Each of the four rows requires the annihilation operator to be moved to its standard position on the right, and the exchange of summation indices in the antisymmetrized interaction then brings the tensor to the form $Y$ or $X$ defined in Eq.~(\ref{eq:TXY}).
In the first row, for example, the product is $V_{rk_4}\an{k_4}\,U^*_{sk_1}\ad{k_1}\,V^*_{s'k_2}\ad{k_2}\,V^*_{r'k_3}\ad{k_3}$.
Moving $\an{k_4}$ to the right past three operators gives a factor $-1$, and the exchange $r\leftrightarrow s$ in the summation gives another factor $-1$ through $\vb_{sr,r's'}=-\vb_{rs,r's'}$, so that the row contributes $+\tfrac14Y$.
The four rows add to $\tfrac14(Y+Y-X-X)=\tfrac12(Y-X)$, so that
\begin{equation}
 \tfrac12\sum_{k_1k_2k_3k_4}(Y-X)_{k_1k_2k_3;k_4}\,\ad{k_1}\ad{k_2}\ad{k_3}\an{k_4}.
\end{equation}

\begin{table}[tb]
\caption{As Table~\ref{tab:31}, for the channels with two quasiparticle creation operators and two annihilation
operators and the order $\ad{k_1}\ad{k_2}\an{k_4}\an{k_3}$.}
\label{tab:22}
\begin{ruledtabular}
\begin{tabular}{lccc}
Operators giving $\alpha^\dagger$ & Sign & Tensor & Net\\
\hline
$a^\dagger_r,a^\dagger_s$ & $+$ & $X^a_{1234}$ & $+X^a$\\
$a_{s'},a_{r'}$ & $+$ & $X^b_{1234}$ & $+X^b$\\
$a^\dagger_r,a_{s'}$ & $-$ & $+X^c_{1234}$ & $-X^c$\\
$a^\dagger_r,a_{r'}$ & $+$ & $-X^c_{1234}$ ($r'\leftrightarrow s'$) & $-X^c$\\
$a^\dagger_s,a_{s'}$ & $+$ & $-X^c_{1234}$ ($r\leftrightarrow s$) & $-X^c$\\
$a^\dagger_s,a_{r'}$ & $-$ & $+X^c_{1234}$ (both) & $-X^c$\\
\end{tabular}
\end{ruledtabular}
\end{table}
The channels in which two of the four operators give creation operators and the other two give annihilation operators are listed in Table~\ref{tab:22}.
Two of the six channels, in which both creation operators or both annihilation operators of $\alpha$ come from $a^\dagger_ra^\dagger_s$ or $a_{s'}a_{r'}$, give the tensors $X^a$ and $X^b$ with sign $+1$.
The four mixed channels, in which each of the pairs $a^\dagger_ra^\dagger_s$ and $a_{s'}a_{r'}$ contributes one creation and one annihilation operator, all give the same net contribution $-X^c$ after the summation indices are relabeled.
The whole group therefore reads
\begin{equation}
 \tfrac14\sum_{k_1k_2k_3k_4}\bigl(X^a+X^b-4X^c\bigr)_{k_1k_2k_3k_4}\,\ad{k_1}\ad{k_2}\an{k_4}\an{k_3}.
\end{equation}

\subsection{Antisymmetrization and the vertices}

Only the part of a coefficient that is antisymmetric in the creation indices and in the annihilation indices contributes to an operator such as
$\ad{k_1}\ad{k_2}\ad{k_3}\ad{k_4}$.  Let $\mathcal A[T]=\frac1{24}\sum_\pi{\rm sgn}(\pi)\,T_{\pi(1234)}$.  Comparing $\tfrac14\sum T\,\ad{}\ad{}\ad{}\ad{}$ with
the definition $\tfrac1{24}\sum H^{40}\ad{}\ad{}\ad{}\ad{}$ gives
$H^{40}=6\,\mathcal A[T]=\tfrac14\sum_\pi{\rm sgn}(\pi)T_{\pi(1234)}$.  Since $T$ is already antisymmetric in $(k_1k_2)$ and in $(k_3k_4)$, the 24 permutations of the indices fall into six classes of four equal terms. Each class is characterized by the pair of indices that occupy the first two positions, and its sign is that of the permutation which brings the indices into this order. One thus obtains Eq.~\eqref{eq:H40}.

For $H^{31}$, comparison of $\tfrac12\sum(Y-X)\ad{}\ad{}\ad{}\an{}$ with $\tfrac16\sum H^{31}\ad{}\ad{}\ad{}\an{}$ gives
$H^{31}=3\,\mathcal A_{123}[Y-X]=\tfrac12\sum_{\pi\in S_3}{\rm sgn}(\pi)(Y-X)_{\pi(123);4}$.  $Y$ is antisymmetric in its last two creation indices and
$X$ in its first two, so each sum has three distinct terms, each twice:
\begin{align}
 H^{31}_{1234}&=\bigl[Y_{123;4}-Y_{213;4}+Y_{312;4}\bigr]\nonumber\\
 &\quad-\bigl[X_{123;4}-X_{132;4}+X_{231;4}\bigr].
\end{align}
Here $Y$ is summed over which index occupies its $U^*$ slot and $X$ over which occupies its $V^*$ slot.  For $H^{22}$, $X^a$ and $X^b$ are already antisymmetric in both
index pairs, and the antisymmetrizer of $-X^c$ over $(12)$ and $(34)$ is
$-\tfrac14\bigl[X^c_{1234}-X^c_{2134}-X^c_{1243}+X^c_{2143}\bigr]$. 
Multiplying by the factor $4$ from Table~\ref{tab:22} gives Eq. \eqref{eq:H22}.

\section{EQUATIONS OF MOTION AND MATRIX BLOCKS}
\label{app:mat}
%=====================================================================

\subsection{Equations of motion in the quasiboson approximation}

Assume $H^{20}=0$, so $H\vac=(E_0+H^{40})\vac$.  Taking $A_a,A^\dagger_b$ with $|a|,|b|\in\{2,4\}$ quasiparticles and expanding
the double commutator we get
\begin{align}
 \bra[A_a,[H,A_b^\dagger]]\vac&=\bra A_aHA^\dagger_b\vac-\bra A_aA^\dagger_bH\vac\nonumber\\
 &\quad-\bra HA^\dagger_bA_a\vac+\bra A^\dagger_bHA_a\vac .
\end{align}
The last two terms vanish ($A_a\vac=0$ and $\bra A^\dagger_b=0$).  In the second, $H\vac=E_0\vac+H^{40}\vac$ and $A_aA^\dagger_b$ changes the
quasiparticle number by $|b|-|a|\ge-2$, so it cannot connect the four-quasiparticle state $H^{40}\vac$ to the vacuum, i.e. the term equals $E_0\delta_{ab}$.  Hence
\begin{equation}
 \Amat_{ab}=\bra A_a(H-E_0)A^\dagger_b\vac=\bra[A_a,[H,A^\dagger_b]]\vac.
\end{equation}
Likewise $\bra[A_a,[H,A_b]]\vac=-\bra A_aA_bH\vac$, so 
\begin{equation}
\Bmat_{ab}=-\bra[A_a,[H,A_b]]\vac. 
\end{equation}
%. these are the forms used in the numerical tests described at the end of this appendix.

Inserting $Q^\dagger=\sum_b(\mathcal{X}_bA^\dagger_b-\mathcal{Y}_bA_b)$ into Eq.~(\ref{eq:EOM}) with $\delta Q=A_c$ we obtain:
\begin{equation}
 \sum_b\bigl(\Amat_{cb}\mathcal{X}_b+\Bmat_{cb}\mathcal{Y}_b\bigr)=\omega \mathcal{X}_c,
\end{equation}
because $\bra[A_c,A^\dagger_b]\vac=\delta_{cb}$ and $\bra[A_c,A_b]\vac=0$.
With $\delta Q=A^\dagger_c$: 
\begin{align}
    \bra[A^\dagger_c,[H,A^\dagger_b]]\vac=&-\bra HA^\dagger_bA^\dagger_c\vac=-\Bmat^*_{cb},\\
    \bra[A^\dagger_c,[H,A_b]]\vac=&\Amat_{bc}=\Amat^*_{cb},\\ 
    \bra[A^\dagger_c,A_b]\vac=&-\delta_{cb},
\end{align}
hence
\begin{equation}
 \sum_b\bigl(\Bmat^*_{cb}\mathcal{X}_b+\Amat^*_{cb}\mathcal{Y}_b\bigr)=-\omega \mathcal{Y}_c .
\end{equation}
Together these are Eq.~(\ref{eq:SQRPA}).  For real symmetric $\Amat$ and $\Bmat$, adding and subtracting the two rows gives
$(\Amat+\Bmat)(\mathcal{X}+\mathcal{Y})=\omega(\mathcal{X}-\mathcal{Y})$ and $(\Amat-\Bmat)(\mathcal{X}-\mathcal{Y})=\omega(\mathcal{X}+\mathcal{Y})$, hence
\begin{equation}
 (\Amat-\Bmat)(\Amat+\Bmat)\,(\mathcal{X}+\mathcal{Y})=\omega^2\,(\mathcal{X}+\mathcal{Y}).
 \label{eq:w2}
\end{equation}
This is the form used for the model calculations ($\omega^2<0$ means an imaginary frequency).

\subsection{Selection rule and elementary matrix elements}

Let $\hat N_q=\sum_k\ad k\an k$.  Then $[\hat N_q,H^{mn}]=(m-n)H^{mn}$ for a vertex with $m$ creation and $n$ annihilation operators, so
$\bra\,\text{($|a|$ qp)}\;H^{mn}\;\text{($|b|$ qp)}\,\vac\neq0$ requires $|a|=|b|+m-n$.  Therefore $\Amat_{11}$ and $\Amat_{22}$ receive $H^{11}$ and $H^{22}$ ($m=n$), $\Amat_{12}$ ($|a|=2$, $|b|=4$) receives
$H^{13}=(H^{31})^\dagger$ only ($m-n=-2$), and $H^{40}$ ($m-n=4$) does not enter $\Amat$ at all.  For the ordered index sets used here
(all indices increasing),
\begin{align}
 \bra A_aA^\dagger_b\vac&=\delta_{ab},\label{eq:E1}\\
 \an d\an c\,\ad{l_1}\ad{l_2}\ad{l_3}\ad{l_4}\vac&=\sum_{m<n}(-1)^{m+n+1}\nonumber\\
 &\quad\times\bigl(\delta_{cl_m}\delta_{dl_n}-\delta_{cl_n}\delta_{dl_m}\bigr)\nonumber\\
 &\quad\times\ad{l_1}\!\cdots\widehat{\ad{l_m}}\cdots\widehat{\ad{l_n}}\cdots\ad{l_4}\vac,\label{eq:E2}\\
 \an{r_3}\an{r_2}\an{r_1}\ad{l_1}\!\cdots\ad{l_4}\vac&=(-1)^d\,\ad{l_d}\vac,\label{eq:E3}
\end{align}
\begin{equation}
 \bra\an{k_4}\an{k_3}\an{k_2}\an{k_1}\ad{k_i}\ad{k_j}\ad{r_1}\ad{r_2}\vac=(-1)^{i+j+1}.\label{eq:E4}
\end{equation}

In Eq.~\eqref{eq:E2} the hats denote omitted operators and $m,n\in\{1,\dots,4\}$. In Eq.~\eqref{eq:E3}, $(r_1,r_2,r_3)$ are the three $l$'s other than $l_d$, in increasing order. In Eq.~\eqref{eq:E4}, $i<j$ and $r_1<r_2$ are the complementary elements of $(k_1k_2k_3k_4)$.

To prove Eq.~\eqref{eq:E2}, we anticommute $\an c$ through $m-1$ creation operators, which gives $(-1)^{m-1}\delta_{cl_m}$. In the second step $\an d$ meets $l_n$ at position $n-1$ for $n>m$ and at position $n$ for $n<m$. After renaming the indices so that $m<n$, this gives Eq.~\eqref{eq:E2}.
Equation~\eqref{eq:E3} follows by moving $\ad{l_d}$ to the end of the product, which requires $4-d$ transpositions and gives the sign $(-1)^{4-d}=(-1)^d$. The three annihilation operators then cancel the remaining three creation operators pairwise. Equation~\eqref{eq:E4} follows from the fact that sorting $(k_i,k_j,r_1,r_2)$ into $(k_1,\dots,k_4)$ requires $(i-1)+(j-2)$ transpositions, from which follows $(-1)^{i+j+1}$.

\subsection{Block \texorpdfstring{$\Bmat$}{B}}

To evaluate $\Bmat_{ab}=\bra A_aA_bH\vac$ we use that, for $H^{20}=0$, $H\vac=(E_0+H^{40})\vac$. The term $E_0\bra A_aA_b\vac$ vanishes because $A_b\vac=0$. Acting with $H^{40}$ on the vacuum gives a four-quasiparticle state, so a nonzero overlap requires $|a|+|b|=4$, which means $|a|=|b|=2$. Therefore $\Bmat_{12}=\Bmat_{21}=\Bmat_{22}=0$.
For $a=(kk')$ and $b=(ll')$, Eq.~\eqref{eq:Hdef} gives $\Bmat_{11}=\bra\an{k'}\an k\an{l'}\an l\,H\vac=H^{40}_{ll'kk'}$, and the exchange of the two index pairs, which is an even permutation of a totally antisymmetric tensor, leads to Eq.~\eqref{eq:B11}. The symmetry $\Bmat_{ab}=\Bmat_{ba}$ follows from the fact that the operators $A_a=\an{k'}\an k$ of the 2qp space commute.

\subsection{Block \texorpdfstring{$\Amat_{11}$}{A11}}

We first evaluate the contribution of $H^{11}=\sum_{mn}e_{mn}\ad m\an n$ to $\Amat_{11}$ for $e_{mn}=E_m\delta_{mn}$. Since $\an n\ad l\ad{l'}\vac=\delta_{nl}\ad{l'}\vac-\delta_{nl'}\ad l\vac$, one has
\begin{equation}
 H^{11}\ad l\ad{l'}\vac=(E_l+E_{l'})\ad l\ad{l'}\vac ,
\end{equation}
and Eq.~\eqref{eq:E1} gives $(E_k+E_{k'})\delta_{kl}\delta_{k'l'}$.

Next we evaluate the contribution of $H^{22}$ to $\Amat_{11}$. The same steps that lead to Eq.~\eqref{eq:E2}, applied to a two-quasiparticle state, give $\an d\an c\ad l\ad{l'}\vac=(\delta_{cl}\delta_{dl'}-\delta_{cl'}\delta_{dl})\vac$, and therefore
\begin{align}
 H^{22}\ad l\ad{l'}\vac&=\tfrac14\sum_{ab}\bigl(H^{22}_{abll'}-H^{22}_{abl'l}\bigr)\ad a\ad b\vac\nonumber\\
 &=\tfrac12\sum_{ab}H^{22}_{abll'}\ad a\ad b\vac .
\end{align}
Projecting on $\bra\an{k'}\an k$ and using $\bra\an{k'}\an k\ad a\ad b\vac=\delta_{ka}\delta_{k'b}-\delta_{kb}\delta_{k'a}$, one obtains $\tfrac12\bigl(H^{22}_{kk'll'}-H^{22}_{k'kll'}\bigr)=H^{22}_{kk'll'}$, which together with the result for $H^{11}$ gives Eq.~\eqref{eq:A11}.

\subsection{Block \texorpdfstring{$\Amat_{22}$}{A22}}

$H^{11}$ gives $\sum_iE_{k_i}\delta_{\{k\}\{l\}}$ by the same argument applied to four quasiparticles.  For $H^{22}$, Eq.~\eqref{eq:E2} shows that $\an d\an c$ removes the pair $(l_m,l_n)$,
$m<n$, with sign $(-1)^{m+n+1}$ and weight $\delta_{cl_m}\delta_{dl_n}-\delta_{cl_n}\delta_{dl_m}$, and the antisymmetry of $H^{22}$ turns the two terms into one:
\begin{align}
 H^{22}\,\ad{l_1}\!\cdots\ad{l_4}\vac&=\tfrac12\sum_{m<n}(-1)^{m+n+1}\nonumber\\
 &\quad\times\sum_{ab}H^{22}_{abl_ml_n}\ad a\ad b\,\ad{r_1}\ad{r_2}\vac,
\end{align}
with $r_1<r_2$ the other two $l$'s.  Projecting with Eq.~\eqref{eq:E4}, a term survives only if $\{a,b\}=\{k_i,k_j\}$ and the two remaining quasiparticles
coincide, $\{r_1,r_2\}=\{k\}\setminus\{k_ik_j\}$, which is the Kronecker symbol in Eq.~\eqref{eq:A22}.  The orderings
$(a,b)=(k_i,k_j)$ and $(k_j,k_i)$ give equal contributions, because both $H^{22}$ and $\ad a\ad b$ change sign under
the exchange, and together they cancel the factor $\tfrac12$.  With the sign $(-1)^{i+j+1}$ of Eq.~\eqref{eq:E4} the
total sign is $(-1)^{m+n+1}(-1)^{i+j+1}=(-1)^{i+j+m+n}$, which gives Eq.~\eqref{eq:A22}.

\subsection{Block \texorpdfstring{$\Amat_{12}$}{A12}}

By the selection rule only $H^{13}=(H^{31})^\dagger=\tfrac16\sum_{abc;e}H^{31*}_{abce}\,\ad e\an c\an b\an a$ contributes.  For
$\ad{l_1}\ad{l_2}\ad{l_3}\ad{l_4}\vac$, Eq.~\eqref{eq:E3} applies to $(a,b,c)$ equal to a permutation $\pi$ of $(r_1,r_2,r_3)$, with an extra factor ${\rm sgn}(\pi)$.
The coefficient $H^{31*}_{abce}$ is antisymmetric in $(abc)$ and supplies the same factor, so the six permutations contribute equally:
\begin{equation}
 H^{13}\ad{l_1}\!\cdots\ad{l_4}\vac=\sum_{d=1}^{4}(-1)^d\sum_eH^{31*}_{r_1r_2r_3e}\;\ad e\ad{l_d}\vac,
\end{equation}
where $(r_1,r_2,r_3)$ are the $l$'s other than $l_d$.  Projecting on $\bra\an{k'}\an k$,
$\bra\an{k'}\an k\ad e\ad{l_d}\vac=\delta_{ke}\delta_{k'l_d}-\delta_{k'e}\delta_{kl_d}$, which gives Eq.~\eqref{eq:A12}.
$\Amat_{21}=\Amat_{12}^\dagger$ since $\Amat$ is Hermitian.

\subsection{Numerical verification of the algebra}
\label{sec:check}

The formulas of Secs.~\ref{sec:HQP}--\ref{sec:red} were verified in a Fock space of $n=6$ fermionic modes,
where all operators are explicit $64\times64$ matrices.  Two tests were done.

In the first test, a random Hamiltonian of the form (\ref{eq:Hdec}) (random $E_k$ and
random antisymmetrized $H^{22},H^{31},H^{40}$) was used with $N_1=15$ two- and
$N_2=15$ four-quasiparticle configurations.  The blocks of $\Amat$ and $\Bmat$
were computed from the commutator definitions $\Amat_{ab}=\bra[A_a,[H,A_b^\dagger]]\vac$,
$\Bmat_{ab}=-\bra[A_a,[H,A_b]]\vac$ and compared with Eqs.~(\ref{eq:A11})--(\ref{eq:A22}).

In the second test, a random complex Bogoliubov transformation and a random complex antisymmetrized $\vb_{rs,r's'}$
with the proper hermiticity were generated.  The $H^{20}$ part of $H$ was removed and $H^{40},H^{31},H^{22}$ were computed from the exact
matrix elements (\ref{eq:Hdef}) and compared with Eqs.~(\ref{eq:H40})--(\ref{eq:H22}).

\begin{table}[tb]
\caption{Largest absolute deviation between the analytic expressions and exact Fock space values.}
\label{tab:check}
\begin{ruledtabular}
\begin{tabular}{lc}
Quantity & max.\ deviation\\
\hline
$\Amat_{11}$, Eq.~(\ref{eq:A11}) & $1.8\times10^{-15}$\\
$\Amat_{12}$, Eq.~(\ref{eq:A12}) & $6.8\times10^{-16}$\\
$\Amat_{22}$, Eq.~(\ref{eq:A22}) & $5.3\times10^{-15}$\\
$\Bmat_{11}$, Eq.~(\ref{eq:B11}) & $3.4\times10^{-16}$\\
$\Bmat_{12},\Bmat_{21},\Bmat_{22}$ & exactly $0$\\
$H^{40}$, Eq.~(\ref{eq:H40}) & $1.5\times10^{-15}$\\
$H^{31}$, Eq.~(\ref{eq:H31}) & $7.3\times10^{-15}$\\
$H^{22}$, Eq.~(\ref{eq:H22}) & $2.7\times10^{-15}$\\
\end{tabular}
\end{ruledtabular}
\end{table}

Table~\ref{tab:check} summarizes the results.  As a third test, one eigenvalue $\omega$ of the full
problem (\ref{eq:SQRPA}) was reinserted into the reduced problem (\ref{eq:eff}). The
reduced problem has an eigenvalue equal to $\omega$ to $\sim10^{-15}$.  Since the random Hamiltonian is not
required to be stable, this eigenvalue was complex, and the test confirms that the
elimination is an exact identity independent of stability.

For the pairing model with $L=6$ and $8$ the same construction was used with $n=2L$ modes, $p$ and $\bar p$ for each
level, which gives $2^{2L}=4096$ and $65\,536$ states.  The particle creation and annihilation operators are explicit
sparse matrices obtained with the Jordan--Wigner representation, and the quasiparticle operators follow from them by
the Bogoliubov transformation.  The quasiparticle vacuum $\vac=\prod_p(u_p+v_pa^\dagger_pa^\dagger_{\bar p})|-\rangle$
is checked to be annihilated by all $\alpha_k$ to $10^{-12}$.  The configuration vectors $A^\dagger_a\vac$ are built
explicitly, and $\Amat$ and $\Bmat$ follow from Eq.~\eqref{eq:ABdef} as matrix elements between these vectors.  The
exact energies with which the QRPA and SQRPA results are compared are obtained by diagonalizing $H$ in the subspace
spanned by the $\binom{L}{N/2}$ states with $N/2$ pairs, $\prod_{p\in\mathcal S}a^\dagger_pa^\dagger_{\bar p}|-\rangle$.
It is verified numerically that $H$ does not lead out of this subspace.

%=====================================================================
\section{ELIMINATION OF THE FOUR-QUASIPARTICLE SPACE, NORMALIZATION AND SUBTRACTION}
\label{app:elim}
%=====================================================================

\subsection{Elimination}

With $\Bmat_{12}=\Bmat_{21}=\Bmat_{22}=0$ the four block rows of Eq.~(\ref{eq:SQRPA}), in the order $(\mathcal{X}_1,\mathcal{X}_2;\mathcal{Y}_1,\mathcal{Y}_2)$, read
\begin{align}
 \Amat_{11}\mathcal{X}_1+\Amat_{12}\mathcal{X}_2+\Bmat_{11}\mathcal{Y}_1&=\omega \mathcal{X}_1,\label{eq:r1}\\
 \Amat_{21}\mathcal{X}_1+\Amat_{22}\mathcal{X}_2&=\omega \mathcal{X}_2,\label{eq:r2}\\
 \Bmat_{11}^*\mathcal{X}_1+\Amat_{11}^*\mathcal{Y}_1+\Amat_{12}^*\mathcal{Y}_2&=-\omega \mathcal{Y}_1,\label{eq:r3}\\
 \Amat_{21}^*\mathcal{Y}_1+\Amat_{22}^*\mathcal{Y}_2&=-\omega \mathcal{Y}_2.\label{eq:r4}
\end{align}
Equations (\ref{eq:r2}) and (\ref{eq:r4}) give
\begin{align}
 \mathcal{X}_2&=(\omega-\Amat_{22})^{-1}\Amat_{21}\mathcal{X}_1,\nonumber\\
 \mathcal{Y}_2&=-(\omega+\Amat_{22}^*)^{-1}\Amat_{21}^*\mathcal{Y}_1,
\end{align}
which is possible when $\omega$ is not an eigenvalue of $\Amat_{22}$ (or $-\Amat_{22}^*$).  Inserting in (\ref{eq:r1}), (\ref{eq:r3}) gives Eq.~(\ref{eq:eff}) with
\begin{align}
 \Sigma(\omega)&=\Amat_{12}(\omega-\Amat_{22})^{-1}\Amat_{21},\nonumber\\
 \tilde\Sigma(\omega)&=-\Amat_{12}^*(\omega+\Amat_{22}^*)^{-1}\Amat_{21}^*.
\end{align}
Conversely, if $(\mathcal{X}_1,\mathcal{Y}_1)$ solves Eq.~(\ref{eq:eff}) at the energy $\omega$, then $(\mathcal{X}_1,\mathcal{X}_2,\mathcal{Y}_1,\mathcal{Y}_2)$ with $\mathcal{X}_2,\mathcal{Y}_2$ as above solves the full problem, so the two
problems have the same eigenvalues (the test in Sec.~\ref{sec:check} confirms this for a random Hamiltonian to $10^{-15}$).

\subsection{Normalization}

The full normalization is $\mathcal{X}_1^\dagger \mathcal{X}_1+\mathcal{X}_2^\dagger \mathcal{X}_2-\mathcal{Y}_1^\dagger \mathcal{Y}_1-\mathcal{Y}_2^\dagger \mathcal{Y}_2=1$.  For real $\omega$ and $\Amat_{21}=\Amat_{12}^\dagger$,
\begin{align}
 \mathcal{X}_2^\dagger \mathcal{X}_2&=\mathcal{X}_1^\dagger\Amat_{12}(\omega-\Amat_{22})^{-2}\Amat_{21}\mathcal{X}_1=-\mathcal{X}_1^\dagger\,\partial_\omega\Sigma\,\mathcal{X}_1,\nonumber\\
 \mathcal{Y}_2^\dagger \mathcal{Y}_2&=\mathcal{Y}_1^\dagger\Amat_{12}^*(\omega+\Amat_{22}^*)^{-2}\Amat_{21}^*\mathcal{Y}_1=\mathcal{Y}_1^\dagger\,\partial_\omega\tilde\Sigma\,\mathcal{Y}_1,
\end{align}
where $\partial_\omega\Sigma=-\Amat_{12}(\omega-\Amat_{22})^{-2}\Amat_{21}$ and $\partial_\omega\tilde\Sigma=\Amat_{12}^*(\omega+\Amat_{22}^*)^{-2}\Amat_{21}^*$.
Hence
\begin{equation}
 \mathcal{X}_1^\dagger(1-\partial_\omega\Sigma)\mathcal{X}_1-\mathcal{Y}_1^\dagger(1+\partial_\omega\tilde\Sigma)\mathcal{Y}_1=1,
\end{equation}
the normalization quoted in Sec.~\ref{sec:red}.  Since $\Amat_{12}$ and $\Amat_{21}=\Amat_{12}^\dagger$ enter as a product with its adjoint, $-\partial_\omega\Sigma$ and $\partial_\omega\tilde\Sigma$ are positive
semidefinite for real $\omega$ outside the spectra of $\pm\Amat_{22}$, i.e., the four-quasiparticle components of $\mathcal{X}$ add to, and those of $\mathcal{Y}$ subtract from, the normalization sum.

\subsection{Subtraction}

At $\omega=0$,
$\Sigma(0)=-\Amat_{12}\Amat_{22}^{-1}\Amat_{21}$ and $\tilde\Sigma(0)=-\Amat_{12}^*(\Amat_{22}^*)^{-1}\Amat_{21}^*$, both independent of $\omega$.  If $\Amat_{22}$ is
positive definite (as in all cases of Sec.~\ref{sec:toy}), $\Sigma(0)$ and $\tilde\Sigma(0)$ are negative semidefinite, i.e.\ the static four-quasiparticle admixture lowers the diagonal
blocks.  The subtracted problem has the blocks
\begin{align}
 \Amat_{11}+\Sigma(\omega)-\Sigma(0)&=\Amat_{11}'+\Sigma(\omega),\nonumber\\
 \Amat_{11}^*+\tilde\Sigma(\omega)-\tilde\Sigma(0)&=\Amat_{11}'^{\,*}+\tilde\Sigma(\omega),\nonumber\\
 \Amat_{11}'&=\Amat_{11}+\Amat_{12}\Amat_{22}^{-1}\Amat_{21}.
\end{align}
Comparing with the elimination above, these are exactly the blocks obtained by eliminating $\mathcal{X}_2,\mathcal{Y}_2$ from the full SQRPA matrix in which $\Amat_{11}$ is replaced by $\Amat_{11}'$ and everything else is unchanged.
This is Eq.~(\ref{eq:A11p}).  Two properties follow.

(i) At $\omega=0$ the effective matrix is
$\bigl(\begin{smallmatrix}\Amat_{11}&\Bmat_{11}\\ \Bmat_{11}^*&\Amat_{11}^*\end{smallmatrix}\bigr)$, the QRPA matrix.

(ii) If $(\mathcal{X}_{\rm sp},\mathcal{Y}_{\rm sp})$ obeys the QRPA equation with $\omega=0$, then by (i) it solves the effective equation at $\omega=0$, with $\mathcal{X}_2=-\Amat_{22}^{-1}\Amat_{21}\mathcal{X}_{\rm sp}$,
$\mathcal{Y}_2=-(\Amat_{22}^*)^{-1}\Amat_{21}^*\mathcal{Y}_{\rm sp}$ it solves the subtracted full problem with $\omega=0$.
In the unsubtracted problem the matrix at $\omega=0$ is the QRPA matrix plus $\Sigma(0)$ and $\tilde\Sigma(0)$ on the diagonal blocks, so in general
the QRPA zero-mode amplitudes are not a solution.  We do not claim a general result for the sign of the resulting shift of $\omega^2$. In the model of Sec.~\ref{sec:toy} it is negative for all couplings.

%=====================================================================
\section{PAIRING MODEL: HFB SOLUTION AND CONFIGURATION SPACES}
\label{app:model}
%=====================================================================

\subsection{Bogoliubov transformation of the model}

The levels $p=1,\dots,L$ hold the time-reversed states $p,\bar p$.  The transformation
\begin{equation}
\begin{aligned}
\alpha_p=u_pa_p-v_pa^\dagger_{\bar p},\\
\alpha_{\bar p}=u_pa_{\bar p}+v_pa^\dagger_p,\\
u_p^2+v_p^2=1,
\end{aligned}
\end{equation}
with real $u_p,v_p$, is of the form \eqref{eq:bog} with $U_{pp}=U_{\bar p\bar p}=u_p$, $V_{\bar pp}=-v_p$, $V_{p\bar p}=v_p$ and all other elements zero.  Conditions
\eqref{eq:canon1} and \eqref{eq:canon2} hold level by level.  The vacuum is the BCS state $\vac=\prod_p(u_p+v_pa^\dagger_pa^\dagger_{\bar p})|-\rangle$, i,e,
$\alpha_p$ and $\alpha_{\bar p}$ annihilate it (Appendix~\ref{app:chain}).

\subsection{HFB energy and equations}

For $\vac$,
\begin{equation}
\langle a^\dagger_pa^\dagger_{\bar p}\rangle=u_pv_p,    
\end{equation} 
and, for different levels, expectation values factorize. Therefore
$\langle a^\dagger_pa^\dagger_{\bar p}a_{\bar q}a_q\rangle=u_pv_pu_qv_q$ for $p\neq q$ and $=\langle n_pn_{\bar p}\rangle=v_p^2$ for $p=q$, so
\begin{equation}
\begin{aligned}
\langle P^\dagger P\rangle~=&~(\sum_pu_pv_p)^2-\sum_pu_p^2v_p^2+\sum_pv_p^2
\\
~=&~(\sum_pu_pv_p)^2+\sum_pv_p^4, 
\end{aligned}
\end{equation}
and
\begin{equation}
 \langle H\rangle=2\sum_p\varepsilon_pv_p^2-G\Bigl[\Bigl(\sum_pu_pv_p\Bigr)^2+\sum_pv_p^4\Bigr].
 \label{eq:EHFB}
\end{equation}
The last term is the Hartree--Fock (self-energy) part and is kept.  In the variation $\vac$ is a trial state with
$u_p=\cos\varphi_p$ and $v_p=\sin\varphi_p$, and the constraint $2\sum_pv_p^2=N$ is imposed with the Lagrange
multiplier $\lambda$.  With the pairing gap and the shifted single-particle energies
\begin{equation}
 \Delta=G\sum_pu_pv_p,\qquad \tilde\varepsilon_p=\varepsilon_p-Gv_p^2-\lambda,
 \label{eq:Deps}
\end{equation}
the derivative
\begin{equation}
 \frac{\partial\langle H-\lambda\hat N\rangle}{\partial\varphi_p}=2\bigl(\tilde\varepsilon_p\sin2\varphi_p-\Delta\cos2\varphi_p\bigr)
 \label{eq:dHdphi}
\end{equation}
vanishes for $\tan2\varphi_p=\Delta/\tilde\varepsilon_p$.  The branch with $\sin2\varphi_p=\Delta/E_p$ and
$\cos2\varphi_p=\tilde\varepsilon_p/E_p$, where $E_p=\sqrt{\tilde\varepsilon_p^2+\Delta^2}$, corresponds to the minimum
and gives
\begin{equation}
 v_p^2=\tfrac12\Bigl(1-\frac{\tilde\varepsilon_p}{E_p}\Bigr).
 \label{eq:gap}
\end{equation}
Since $\Delta$ and $\tilde\varepsilon_p$ depend on the occupations, Eqs.~\eqref{eq:Deps} and \eqref{eq:gap} are solved
self-consistently together with the particle-number constraint.  The solution defines the HFB vacuum $\vac$ used in
the rest of the paper.  Evaluated with the general formulas of Appendix~\ref{app:vert}, the quasiparticle energy is
$E_p$.  It agrees with the Fock-space matrix element $\bra\an p H'\ad p\vac-E_0$, where $E_0=\bra H'\vac$, to
$2\times10^{-15}$, and Eq.~\eqref{eq:EHFB} agrees with the Fock-space value of $\bra H\vac$ to the same accuracy.

In the limit $\Delta\to0$ the occupations become $v_p^2=1$ for $p\le N/2$ and $v_p^2=0$ otherwise, and
$E_p\to|\tilde\varepsilon_p|$.  Dividing $\Delta=G\sum_pu_pv_p=G\sum_p\Delta/(2E_p)$ by $\Delta$ gives the condition for
the critical coupling,
\begin{equation}
 1=\frac{G_c}{2}\sum_p\frac{1}{|\tilde\varepsilon_p|},\qquad
 \tilde\varepsilon_p=\begin{cases}\varepsilon_p-G_c-\lambda, & p\le N/2,\\ \varepsilon_p-\lambda, & p>N/2,\end{cases}
 \label{eq:Gc}
\end{equation}
where the shift $-G_c$ is the Hartree--Fock term.  The chemical potential follows from the particle-number condition
to second order in $\Delta$, $\sum_{p>N/2}\tilde\varepsilon_p^{-2}=\sum_{p\le N/2}\tilde\varepsilon_p^{-2}$.

\subsection{Numerical solution}

Starting from $v_p^2=N/2L$, the equations are iterated with $\lambda$ obtained from $\sum_p2v_p^2=N$ by a bracketing root finder
at each step, and the update $v^2\to\tfrac12v^2+\tfrac12v^2_{\rm new}$ until $\max_p|\Delta v_p^2|<10^{-14}$.  The vacuum is a minimum of the energy (checked by a constrained
minimization over $\{v_p\}$).  A solution is called paired if $\Delta>10^{-4}d$. Below the critical coupling the iteration converges to $\Delta\to0$.

\subsection{Configuration spaces and blocking}

The Fock space of one level splits into the even subspace $\{|-\rangle,\;a^\dagger_pa^\dagger_{\bar p}|-\rangle\}$ and the odd subspace
$\{a^\dagger_p|-\rangle,a^\dagger_{\bar p}|-\rangle\}$.  The operators $\alpha_p,\alpha_{\bar p}$ map each even state to an odd one and conversely, so $\alpha^\dagger_p\alpha^\dagger_{\bar p}$ preserves parity on a level
(it acts within the even subspace, see Appendix~\ref{app:chain}) while a single $\alpha^\dagger_p$ changes it.  The operator $\hat{n}_p-\hat{n}_{\bar p}$ commutes with $H$ for every level $p$ (see Appendix~\ref{app:exact}), so $H$ does not connect
configurations in which a given level is odd to those in which it is even.  Consequently the configurations from Eq. \eqref{ApApq} form a closed set for the matrices $\Amat$ and $\Bmat$. In other words, the elements $\Amat_{ab}$ and $\Bmat_{ab}$ involve only even levels, and the
couplings to configurations with unpaired quasiparticles (e.g.\ $\alpha^\dagger_p\alpha^\dagger_q$ with $p\neq q$) vanish.  This was also checked numerically for $L=6$, $G=0.8d$, in which of total $561$ two- and
four-quasiparticle configurations, $21$ have all quasiparticles paired, and the largest matrix element of $H'$ between these and the rest is exactly $0$.  For example, the sizes of $\Amat$ are $L+L(L-1)/2=21,\,36,\,210,$ and $406$ for $L=6,8,20$ and $28$, respectively.

%=====================================================================
\section{CHAIN REPRESENTATION FOR LARGE \texorpdfstring{$L$}{L}}
\label{app:chain}
%=====================================================================

\subsection{Two-state algebra on one level}

\begin{figure}[tb]
  \centering
  \includegraphics[width=\columnwidth]{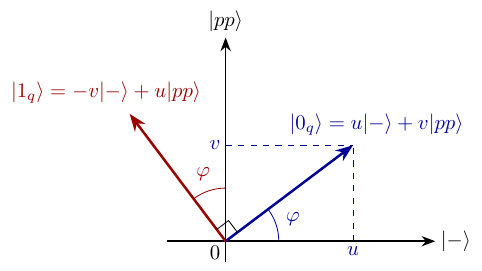}
  \caption{Single-level basis vectors in the plane spanned by the empty-level
  state $|-\rangle$ and the filled-level state
  $|pp\rangle=a^\dagger_pa^\dagger_{\bar p}|-\rangle$.
  The quasiparticle basis $\{|0_q\rangle,|1_q\rangle\}$ is obtained by a
  rotation through the angle $\varphi$.}
  \label{fig:basis}
\end{figure}

On level $p$ the even subspace has the basis $|-\rangle$, $|pp\rangle\equiv a^\dagger_pa^\dagger_{\bar p}|-\rangle$.  The local quasiparticle vacuum and the pair state are
\begin{align}
 |0_q\rangle&=u|-\rangle+v|pp\rangle,\\
 |1_q\rangle&\equiv\alpha^\dagger_p\alpha^\dagger_{\bar p}|0_q\rangle=-v|-\rangle+u|pp\rangle.
 \label{eq:levelstates}
\end{align}
Both states follow from the action of the quasiparticle operators of the level on $|-\rangle$ and $|pp\rangle$.  The
annihilation operator $\alpha_p$ annihilate $|0_q\rangle$,
\begin{equation}
\begin{aligned}
\alpha_p|0_q\rangle=&(ua_p-va^\dagger_{\bar p})(u+va^\dagger_pa^\dagger_{\bar p})|-\rangle\\
=&uv\,a^\dagger_{\bar p}|-\rangle-vu\,a^\dagger_{\bar p}|-\rangle=0,     
\end{aligned}
\end{equation}
and similarly $\alpha_{\bar p}|0_q\rangle=0$. For $\alpha^\dagger_{\bar p}|0_q\rangle$
\begin{equation}
\begin{aligned}
\alpha^\dagger_{\bar p}|0_q\rangle=&(u a^\dagger_{\bar p}+va_p)(u+va^\dagger_pa^\dagger_{\bar p})|-\rangle
\\
=&(u^2+v^2)a^\dagger_{\bar p}|-\rangle=a^\dagger_{\bar p}|-\rangle,    
\end{aligned}
\end{equation}
so the
pair state is obtained with two creation operators,
\begin{equation}
\alpha^\dagger_pa^\dagger_{\bar p}|-\rangle=(ua^\dagger_p-va_{\bar p})a^\dagger_{\bar p}|-\rangle=u|pp\rangle-v|-\rangle.    
\end{equation} 
The two states ($|0_q\rangle$ and $|1_q\rangle$) can be obtained from $|-\rangle$ and $|pp\rangle$ by a rotation through the angle $\varphi_p$ of
Appendix~\ref{app:model}, with $u=\cos\varphi_p$ and $v=\sin\varphi_p$, as shown in Fig.~\ref{fig:basis}.

In the basis $|0_q\rangle$, $|1_q\rangle$ each level is a two-state system, and all operators that act on the even
subspace of the level can be written with the three operators
\begin{equation}
 \sigma^+=|1_q\rangle\langle0_q|,\quad\sigma^-=|0_q\rangle\langle1_q|,\quad n=|1_q\rangle\langle1_q| .
\end{equation}
The operator $\sigma^+$ creates a quasiparticle pair on the level, $\sigma^-$ removes it, and $n$ counts it ($n=0$ or
$1$).  They have the algebra of the Pauli raising, lowering and projection operators of a spin $\tfrac12$.  Since
$\alpha^\dagger_p\alpha^\dagger_{\bar p}$ maps $|0_q\rangle$ to $|1_q\rangle$ and $|1_q\rangle$ to zero, the pair
creation operator of the SQRPA is $A^\dagger_p=\alpha^\dagger_p\alpha^\dagger_{\bar p}=\sigma^+_p$ on the even
subspace.  Operators on different levels contain an even number of fermion operators and therefore commute.

The inverse rotation, through $-\varphi_p$, expresses the particle basis in terms of the quasiparticle basis,
$|-\rangle=u|0_q\rangle-v|1_q\rangle$ and $|pp\rangle=v|0_q\rangle+u|1_q\rangle$.  On the even subspace the pair creation
operator and the particle number of the level are $a^\dagger_pa^\dagger_{\bar p}=|pp\rangle\langle-|$ and
$\hat N_p=2|pp\rangle\langle pp|$.  Inserting the inverse rotation and using $|0_q\rangle\langle0_q|=1-n$ gives
\begin{align}
 a^\dagger_pa^\dagger_{\bar p}&=|pp\rangle\langle-|\nonumber\\
 &=u^2\sigma^+-v^2\sigma^-+uv(1-2n),\label{eq:chainP}\\
 \hat N_p&\equiv a^\dagger_pa_p+a^\dagger_{\bar p}a_{\bar p}=2|pp\rangle\langle pp|\nonumber\\
 &=2v^2+2(u^2-v^2)n+2uv(\sigma^++\sigma^-).\label{eq:chainN}
\end{align}
Both identities were verified numerically on the four-dimensional single-level space (deviation $10^{-16}$).

\subsection{The Hamiltonian as a system of two-state levels}

In the space of paired states each level $p$ has only the two states $|0_q\rangle_p$ and $|1_q\rangle_p$, and the
states of the whole system are products of the states of the individual levels.  This space has $2^L$ states,
compared with $2^{2L}$ in the Fock space, where each level has four states.  A product state is labelled by the set
$S$ of levels that carry a quasiparticle pair ($n_p=1$),
\begin{equation}
 |S\rangle=\prod_{p\in S}\sigma^+_p\vac,\qquad \vac=\prod_p|0_q\rangle_p ,
\end{equation}
and $|S|$ is its number of quasiparticle pairs.  A level carries at most one pair, so the two-quasiparticle
configurations of the SQRPA have one level with a pair ($|S|=1$), and the four-quasiparticle configurations have two
different levels with a pair ($|S|=2$).

The Hamiltonian is $H'=H-\lambda\hat N=\sum_p(\varepsilon_p-\lambda)\hat N_p-GP^\dagger P$ with
$P^\dagger=\sum_pa^\dagger_pa^\dagger_{\bar p}$.  By Eqs.~\eqref{eq:chainP} and \eqref{eq:chainN} the first term is a
sum of operators on single levels, and $P^\dagger P=\sum_{p,q}(a^\dagger_pa^\dagger_{\bar p})(a_{\bar q}a_q)$ is a
sum of products of operators on two levels.  $H'$ is therefore a matrix in the space of the $2^L$ product states that
can be built without the Fock space.  Every pair of levels is coupled with the same strength $G$.  The name chain
representation refers only to the structure of the space as a string of $L$ two-state systems.

\subsection{Truncation}

$H'$ is a sum of single-level operators and of $P^\dagger P=\sum_{pq}(a^\dagger_pa^\dagger_{\bar p})(a_{\bar q}a_q)$, whose factors change the number of excitations by $-1,0,+1$ each
(Eq.~\eqref{eq:chainP}, $P=(P^\dagger)^\dagger$).  Computing $\langle a|P^\dagger P|b\rangle=\sum_c\langle a|P^\dagger|c\rangle\langle c|P|b\rangle$ with $a$ and $b$ having at most two excitations, the intermediate state $c$ differs from $b$ by at most one excitation, so $c$ has at most
three.  The truncation of the chain to at most three excitations is therefore exact for these matrix elements.  The truncated basis has
$1+L+\binom L2+\binom L3$ states, e.g.\ $3683$ for $L=28$, and $H'$ is a sparse matrix in it.

\section{EXACT SOLUTION IN THE SENIORITY-ZERO SPACE}
\label{app:exact}
%=====================================================================

\subsection{Conservation of seniority}

Let $s_p=n_p-n_{\bar p}$.  The operators $a^\dagger_qa^\dagger_{\bar q}$ and $a_{\bar q}a_q$ change $n_q$ and
$n_{\bar q}$ by the same amount, so they commute with every $s_p$, and $s_p$ also commutes with $\hat N$.  Hence
$[H,s_p]=0$ for every level $p$, and $H=\sum_p\varepsilon_p\hat N_p-GP^\dagger P$ is block diagonal in the set
$\{s_p\}$.  The seniority is the number of levels with $s_p\neq0$, which are the singly occupied (blocked) levels.  A
pair $(p,\bar p)$ can be neither created nor removed on such a level.

\subsection{The seniority-zero space}

For $N$ particles with no singly occupied level, all $s_p=0$, a basis is
\begin{equation}
 |\mathcal S\rangle=\prod_{p\in\mathcal S}a^\dagger_pa^\dagger_{\bar p}\,|-\rangle,\qquad
 \mathcal S\subset\{1,\dots,L\},\quad |\mathcal S|=N/2,
\end{equation}
where $\mathcal S$ is the set of doubly occupied levels.  For $N=L$ its dimension $\binom{L}{N/2}$ is $20$, $70$,
$184\,756$ and $40\,116\,600$ for $L=6$, $8$, $20$ and $28$, compared with $2^{2L}=4096$ and $65\,536$ states of the
Fock space for $L=6$ and $8$.  Pair operators on different levels commute, so no fermionic signs appear.  In
$P^\dagger P|\mathcal S\rangle=\sum_{pq}a^\dagger_pa^\dagger_{\bar p}a_{\bar q}a_q|\mathcal S\rangle$ the terms with
$p=q\in\mathcal S$ give $N/2$ times $|\mathcal S\rangle$, and the terms with $q\in\mathcal S$ and $p\notin\mathcal S$
move the pair from $q$ to $p$ with coefficient $+1$.  All other terms vanish, and the nonzero matrix elements are
\begin{align}
 \langle\mathcal S|H|\mathcal S\rangle&=2\sum_{p\in\mathcal S}\varepsilon_p-G\,\tfrac N2,\label{eq:sen00}\\
 \langle\mathcal S'|H|\mathcal S\rangle&=-G\quad(\mathcal S'\ \text{obtained from}\ \mathcal S\ \text{by moving one pair}).\label{eq:sen01}
\end{align}
The exact energies are the eigenvalues of this sparse symmetric matrix, and the excitation energies are their
differences from the lowest eigenvalue. The model corresponds to the pairing Hamiltonian with constant strength that was solved exactly by
Richardson~\cite{Richardson1963,RichardsonSherman1964} (see also Ref.~\cite{Dukelsky2004} for a review).  Here the matrix is
diagonalized directly, and the Richardson equations are not used.

\subsection{Relation to the QRPA configurations and checks}

The configurations $A^\dagger_p\vac$ and $A^\dagger_{pq}\vac$ of Appendix~\ref{app:model} are products of the pair
operators $\alpha^\dagger_p\alpha^\dagger_{\bar p}$ acting on the BCS state and have all $s_p=0$.  They therefore lie
in the seniority-zero space, and the QRPA and SQRPA excitations of Sec.~\ref{sec:toy} are compared with the lowest
excitation in this space.  States with broken pairs, with $s_p\neq0$ on some levels and seniority two or higher, are
not part of the comparison.  Depending on $G$ the lowest of them lies below or above the first seniority-zero
excitation.  For $L=8$ it lies at $2.310\,d$ compared with $2.592\,d$ for $G=0.5\,d$, and at $6.010\,d$ compared with
$5.949\,d$ for $G=d$.

The matrices \eqref{eq:sen00} and \eqref{eq:sen01} were diagonalized independently of the Fock-space code.  The
ground-state energies and the four lowest excitation energies coincide with the Fock-space results at all couplings of
Fig.~\ref{fig:toy} for $L=6$ and $8$.  For $L=N=6$ and $G=d$ we obtain $E_0=-0.18916\,d$ and $\omega_1=4.5572\,d$, as
in Table~\ref{tab:toy}.  For $L=N=20$ and $G=d$, with $184\,756$ states and a sparse (Lanczos) eigensolver, we obtain
$E_0=47.3982\,d$ and $\omega_1=16.2860\,d$ in about $25$\,s.  For $L=28$ the dimension is too large.
\section{NUMBER-PROJECTED MATRIX ELEMENTS IN THE PAIRING MODEL}
\label{app:proj}

The quasiparticle vacuum $\vac$ and the configurations built on it are not eigenstates of the particle number.  In this
appendix we derive the matrix elements of the norm and of the Hamiltonian between number-projected configurations of
the chain representation (Appendix~\ref{app:chain}).  They are exact, require no Fock-space construction, and are
evaluated with $L+1$ gauge angles.  The same matrices are used for two different purposes, the projected energies of
the QRPA, SQRPA and subtracted SQRPA roots and the projected configuration interaction of Sec.~\ref{sec:projth}.  They
do not define a projected SQRPA.

\subsection{Projector and configurations}

In the seniority-zero space the particle number is $\hat N=2\hat n$ with
\begin{equation}
 \hat n=\sum_{p}\nu_p,\qquad \nu_p=b^\dagger_pb_p,\qquad b^\dagger_p=a^\dagger_pa^\dagger_{\bar p},
\end{equation}
so that $\nu_p\in\{0,1\}$ counts the pairs on level $p$.  The projector on $N=2M$ particles is
\begin{equation}
 \hat P^{N}=\frac{1}{2\pi}\int_0^{2\pi}\!d\phi\,e^{i\phi(\hat N-N)}
 =\frac{1}{2\pi i}\oint\frac{dz}{z}\,z^{\hat n-M},
 \label{eq:proj}
\end{equation}
with $z=e^{2i\phi}$.
The gauge operator factorizes over the levels, $z^{\hat n}=\prod_{p}z^{\nu_p}$, and in the basis
$\{|-\rangle_p,|pp\rangle_p\}$ of level $p$ it is $\mathrm{diag}(1,z)$.

The configurations $|S\rangle$ are those of Appendix~\ref{app:chain}, labelled by the set $S$ of levels that carry a
quasiparticle pair, and we write $s_p=1$ for $p\in S$ and $s_p=0$ otherwise.  On level $p$ the state is $|0_q\rangle$
for $s_p=0$ and $|1_q\rangle$ for $s_p=1$ (see Eq.~\eqref{eq:levelstates}), with $u=u_p$ and $v=v_p$.

\subsection{Norm and Hamiltonian kernels}

The norm kernel is a product of level factors,
\begin{equation}
 \mathcal N_{S'S}(z)=\langle S'|z^{\hat n}|S\rangle=\prod_{p}r_p(s'_p,s_p;z),
 \label{eq:normkernel}
\end{equation}
where, from Eq.~\eqref{eq:levelstates}, the $2\times2$ matrix $r_p$ with rows $s'_p$ and columns $s_p$ is
\begin{equation}
 r_p(z)=\begin{pmatrix}u_p^2+v_p^2z & u_pv_p(z-1)\\ u_pv_p(z-1) & v_p^2+u_p^2z\end{pmatrix}.
\end{equation}

In the seniority-zero space $\sum_{p}\varepsilon_p(n_p+n_{\bar p})=2\sum_{p}\varepsilon_p\nu_p$ and
$P^\dagger P=\sum_{p}\nu_p+\sum_{p\neq q}b^\dagger_pb_q$, so that
\begin{equation}
 H=\sum_{p}(2\varepsilon_p-G)\,\nu_p-G\sum_{p\neq q}b^\dagger_pb_q .
 \label{eq:Hsen}
\end{equation}
Because $[H,\hat n]=0$ and $\hat P^N$ is a Hermitian projector, $\langle S'|\hat P^NH\hat P^N|S\rangle=\langle S'|H\hat P^N|S\rangle$,
and only the kernel $\langle S'|Hz^{\hat n}|S\rangle$ is needed.  Each term of Eq.~\eqref{eq:Hsen} acts on one or two
levels, and the kernel follows by replacing the corresponding factors of Eq.~\eqref{eq:normkernel},
\begin{align}
 \mathcal H_{S'S}(z)={}&\sum_{p}(2\varepsilon_p-G)\,t^{\nu}_p(s'_p,s_p;z)\prod_{q\neq p}r_q(s'_q,s_q;z)\nonumber\\
 -G&\sum_{p\neq q}t^{+}_p(s'_p,s_p)\,t^{-}_q(s'_q,s_q;z)\prod_{k\neq p,q}r_k(s'_k,s_k;z),
 \label{eq:Hkernel}
\end{align}
with the level matrices of $\nu_pz^{\nu_p}$, $b^\dagger_pz^{\nu_p}$ and $b_pz^{\nu_p}$,
\begin{align}
 t^{\nu}_p&=z\begin{pmatrix}v_p^2&u_pv_p\\u_pv_p&u_p^2\end{pmatrix},\qquad
 t^{+}_p=\begin{pmatrix}u_pv_p&-v_p^2\\u_p^2&-u_pv_p\end{pmatrix},\nonumber\\
 t^{-}_p&=z\begin{pmatrix}u_pv_p&u_p^2\\-v_p^2&-u_pv_p\end{pmatrix}.
\end{align}
The matrix $t^+_p$ is the representation of $b^\dagger_p=u_p^2\sigma^+_p-v_p^2\sigma^-_p+u_pv_p(1-2n_p)$ given in
Appendix~\ref{app:chain}, and $t^-_p=z\,(t^+_p)^{T}$.

\subsection{Exact projection with a finite number of gauge angles}

Every entry of $r_p$, $t^\nu_p$, $t^\pm_p$ is a polynomial of degree at most one in $z$, so $\mathcal N_{S'S}$ and
$\mathcal H_{S'S}$ are polynomials of degree at most $L$.  By Eq.~\eqref{eq:proj} the projected matrices are their
coefficients of $z^M$,
\begin{equation}
 \mathsf N^{P}_{S'S}=[z^M]\,\mathcal N_{S'S}(z),\qquad \mathsf H^{P}_{S'S}=[z^M]\,\mathcal H_{S'S}(z),
\end{equation}
and the coefficient is obtained exactly from $K=L+1$ points on the unit circle,
\begin{equation}
 [z^M]f=\frac1K\sum_{k=0}^{K-1}z_k^{-M}f(z_k),\qquad z_k=e^{2\pi ik/K},
\end{equation}
which corresponds to the gauge angles $\phi_k=\pi k/K$.
For the vacuum, $S=\emptyset$, denoted by the index $0$ as in Sec.~\ref{sec:projth}, so $\mathsf N^P_{00}$ is the
probability of $N$ particles in $\vac$,
\begin{equation}
 \mathsf N^{P}_{00}=\sum_{|S|=M}\ \prod_{p\in S}v_p^2\prod_{p\notin S}u_p^2 .
\end{equation}
For $L=N=8$ it equals $0.572$ at $G/d=0.5$ and $0.408$ at $G/d=0.7$.

\subsection{Use of the projected matrices}

For the projected energies $\omega^P_\nu$ (see Eq.~\eqref{eq:wP}) of the QRPA, SQRPA and SSQRPA roots, the coefficients
$c^\nu_S$ are the forward amplitudes $X^\nu_S$ of the root on the configurations with one and two pair excitations,
and the vacuum coefficient $c^\nu_0$ follows from Eq.~\eqref{eq:nuP}.  Each root is projected separately, and
the roots are neither mixed nor recomputed.  For the projected configuration interaction, Eq.~\eqref{eq:PCI}, the space
$\mathcal V_n$ contains the vacuum and all configurations with at most $n$ pair excitations, and the matrices are
diagonalized anew without reference to any RPA solution.

The norm matrix of the configuration interaction is singular.  Since $\hat P^N(\hat N-N)=0$, every vector
$(\hat N-N)|S\rangle$ that lies in $\mathcal V_n$ is a null vector.  For $n=1$ this is $(\hat N-N)\vac$ alone, the
Goldstone direction, and for $n=2$ these are the $L+1$ vectors with $|S|\le1$.  Numerically, the null space has exactly
this dimension for $L=8$ and $16$.

\subsection{Numerical check}

For $L=8$ ($n\le2$) and $L=16$ ($n\le1$) the matrices $\mathsf N^P$ and $\mathsf H^P$ agree with a direct construction,
in which every configuration is rotated to the pair basis and restricted to the $N$-particle sector, to
$4\times10^{-14}$.  Table~\ref{tab:pcm} compares the lowest excitation energies of Eq.~\eqref{eq:PCI} with the exact
ones.

\begin{table}[t]
\caption{Exact ground-state energy $E_0$ and lowest seniority-zero excitation energies (in units of $d$) compared with
the projected configuration interaction, Eq.~\eqref{eq:PCI}, in $\mathcal V_1$ and $\mathcal V_2$.  Degenerate exact levels
are listed once per state.}
\label{tab:pcm}
\begin{ruledtabular}
\begin{tabular}{lllcccccc}
$L$ & $G/d$ & & $E_0$ & $\omega_1$ & $\omega_2$ & $\omega_3$ & $\omega_4$ & $\omega_5$\\
\hline
8  & 0.5 & exact          & 8.8892  & 2.592 & 4.557 & 4.557 & 6.542 & 6.542\\
   &     & $\mathcal V_1$ & 8.9542  & 2.699 & 4.674 & 4.844 & 6.673 & 6.778\\
   &     & $\mathcal V_2$ & 8.8912  & 2.596 & 4.562 & 4.564 & 6.544 & 6.551\\
8  & 0.7 & exact          & 6.6910  & 3.635 & 5.418 & 5.418 & 7.302 & 7.302\\
   &     & $\mathcal V_1$ & 6.7510  & 3.609 & 5.470 & 5.581 & 7.348 & 7.413\\
   &     & $\mathcal V_2$ & 6.6920  & 3.642 & 5.432 & 5.432 & 7.312 & 7.325\\
16 & 0.5 & exact          & 48.4918 & 3.715 & 5.396 & 5.396 & 7.205 & 7.205\\
   &     & $\mathcal V_1$ & 48.6834 & 3.658 & 5.488 & 5.644 & 7.343 & 7.452\\
   &     & $\mathcal V_2$ & 48.5005 & 3.763 & 5.488 & 5.494 & 7.286 & 7.322\\
16 & 0.7 & exact          & 41.4299 & 7.040 & 8.044 & 8.044 & 9.446 & 9.446\\
   &     & $\mathcal V_1$ & 41.5740 & 7.332 & 8.090 & 8.197 & 9.484 & 9.568\\
   &     & $\mathcal V_2$ & 41.4324 & 7.285 & 8.185 & 8.252 & 9.558 & 9.616\\
\end{tabular}
\end{ruledtabular}
\end{table}

The projected configuration interaction with two pair excitations reproduces the exact ground-state energy to
$2\times10^{-3}d$ for $L=8$ and to $9\times10^{-3}d$ for $L=16$.  The excitation energies are within $0.05\,d$ for
$L=8$ and within $0.25\,d$ for $L=16$.

\bibliographystyle{apsrev4-2}
\bibliography{references}

\end{document}